\documentclass[twocolumn]{aastex63}
\usepackage{enumitem}
\usepackage{tikz}

\def \mathbi#1{\textbf{\em #1}}

\shorttitle{Disc tearing}
\shortauthors{A. Raj et.~al.}
\begin{document}
\title{Disc tearing: numerical investigation of warped disc instability}

\author{A.~Raj}
\affiliation{School of Physics and Astronomy, University of Leicester, Leicester, LE1 7RH, UK}
\author[0000-0002-2137-4146]{C.~J.~Nixon}
\affiliation{School of Physics and Astronomy, University of Leicester, Leicester, LE1 7RH, UK}
\author{S.~Do\u{g}an}
\affiliation{Department of Astronomy and Space Sciences, University of Ege, Bornova, 35100, ${\dot {\rm I}}$zmir, Turkey}

\email{cjn@leicester.ac.uk}

\begin{abstract}
We present numerical simulations of misaligned discs around a spinning black hole covering a range of parameters. Previous simulations have shown that discs that are strongly warped by a forced precession---in this case the Lense-Thirring effect from the spinning black hole---can break apart into discrete discs or rings that can behave quasi-independently for short timescales. With the simulations we present here, we confirm that thin and highly inclined discs are more susceptible to disc tearing than thicker or low inclination discs, and we show that lower values of the disc viscosity parameter lead to instability at lower warp amplitudes. This is consistent with detailed stability analysis of the warped disc equations. We find that the growth rates of the instability seen in the numerical simulations are similar across a broad range of parameters, and are of the same order as the predicted growth rates. However, we did not find the expected trend of growth rates with viscosity parameter. This may indicate that the growth rates are affected by numerical resolution, or that the wavelength of the fastest growing mode is a function of local disc parameters. Finally, we also find that disc tearing can occur for discs with a viscosity parameter that is higher than predicted by a local stability analysis of the warped disc equations. In this case, the instability manifests differently producing large changes in the disc tilt locally in the disc, rather than the large changes in disc twist that typically occur in lower viscosity discs.
  
\end{abstract}


\section{Introduction}
Accretion discs are generally considered to orbit their central accretor in a smooth sequence of planar and circular orbits. However, there is a substantial body of observational evidence that shows that real discs are more complicated. For example, warped discs are used to model the X-ray binary Her X-1 \citep[e.g.\@][]{Scott:2000aa}, water masers reveal discs warps around supermassive black holes \citep[e.g.\@][]{Miyoshi:1995aa,Greenhill:1995aa}, and the recent spatially resolved observations of protoplanetary discs have directly revealed complex disc structures including disc warps \citep[e.g.][]{Casassus:2015aa}. Very recently it has been suggested that a misalignment between the gas and black hole angular momentum may be required to resolve discrepancies in the mass measurements for the supermassive black hole in M87 that was imaged by the Event Horizon Telescope \citep{Jeter:2020aa}. In different systems warps may form in different ways. For example, warps can form during the formation of an accretion disc \citep[e.g.\@ in chaotic star forming regions;][]{Bate:2010aa,Bate:2018aa}, or the disc may become warped if the orbits are forced to precess in a radially differential manner by the spin of the central object \citep[e.g.\@ a spinning black hole;][]{Bardeen:1975aa} or the gravitational field of a binary system \citep[e.g.\@][]{Papaloizou:1995ab,Larwood:1997aa}.

To a first approximation warps can propagate through a disc in one of two ways dependent on the relative magnitudes of the \cite{Shakura:1973aa} dimensionless viscosity parameter, $\alpha$, and the disc angular semi-thickness, $H/R$ \citep{Papaloizou:1983aa}. For low viscosity discs, where $\alpha < H/R$, warps propagate via bending waves. For high viscosity discs, where $\alpha > H/R$, warps propagate following a diffusion equation. In this work we are principally interested in the diffusive case, although we do present some simulations with $\alpha \lesssim H/R$. For a review of the basics of warped disc dynamics see \cite{Nixon:2016aa}.

Recently it has been possible to explore detailed dynamics of warped discs with numerical hydrodynamical simulations. There is now a broad range of simulations covering a variety of astrophysical setups, including; (1) isolated warped discs \citep[e.g.\@][]{Nelson:1999aa,Lodato:2010aa}, (2) warped discs in binary systems \citep[e.g.\@][]{Larwood:1996aa,Larwood:1997aa,Fragner:2010aa}, and (3) warped discs around spinning black holes \citep[e.g.\@][]{Nelson:2000aa,Fragile:2007aa}. A recurring feature of the warped discs in simulations is that they can `break' or `tear' \citep[e.g.][]{Nixon:2012ad}, and this typically occurs when the simulation parameters allow for large enough warp amplitudes (either present at the start of the simulation, or formed over time) at sufficiently low viscosity.

The warp amplitude, $\psi$, is given by\footnote{Note that here, for clarity, we refer to the warp amplitude by the symbol $\psi$. This is different to some previous works which refer to the warp amplitude as $\left|\psi\right|$ due to the use of complex variables in \cite{Ogilvie:1999aa} in which the warp amplitude was the magnitude, $\left|\psi\right|$, of a complex variable $\psi$ (see equations 44-45 of \citealt{Ogilvie:1999aa}). Here we take $\psi$ to be the warp amplitude.}
\begin{equation}
  \psi(R,t) = R \left|\frac{\partial \mathbi{l}}{\partial R}\right|\,,
\end{equation}
where $R$ is the orbital radius, and $\mathbi{l}(R,t)$ is the unit `tilt' vector pointing in the direction of the disc angular momentum vector. The unit tilt vector is usually written as $\mathbi{l} = (\cos\gamma\sin\beta,\sin\gamma\sin\beta,\cos\beta)$, where $\beta(R,t)$ is the disc `tilt' and $\gamma(R,t)$ is the disc `twist'. From this we can see that discs can be warped in distinct ways. For example, a disc may achieve a warp ($\psi > 0$) if $\beta$ varies with radius or if $\gamma$ varies with radius (and $\beta \ne 0$). We illustrate these two cases in Fig~\ref{Fig0}. As we shall see below, a large warp amplitude can be caused by a sharp variation in $\beta$ or a sharp variation in $\gamma$. In general a warped disc has both $\beta$ and $\gamma$ varying with radius. Locally, on scales $\sim H$, the disc doesn't know whether misaligned neighbouring rings are tilted or twisted with respect to each other, and thus the stability of the disc (as determined by a local stability analysis) is determined by $\psi$ and not $\beta$ or $\gamma$. However, as we shall see below the global behaviour of the disc can result in the local instability manifesting as either a sharp change in the disc twist (\citealt{Nixon:2012ad}) or a sharp change in the disc tilt (see Section~\ref{alpha0.3}).

\begin{figure*}
  \includegraphics[width=0.47\textwidth]{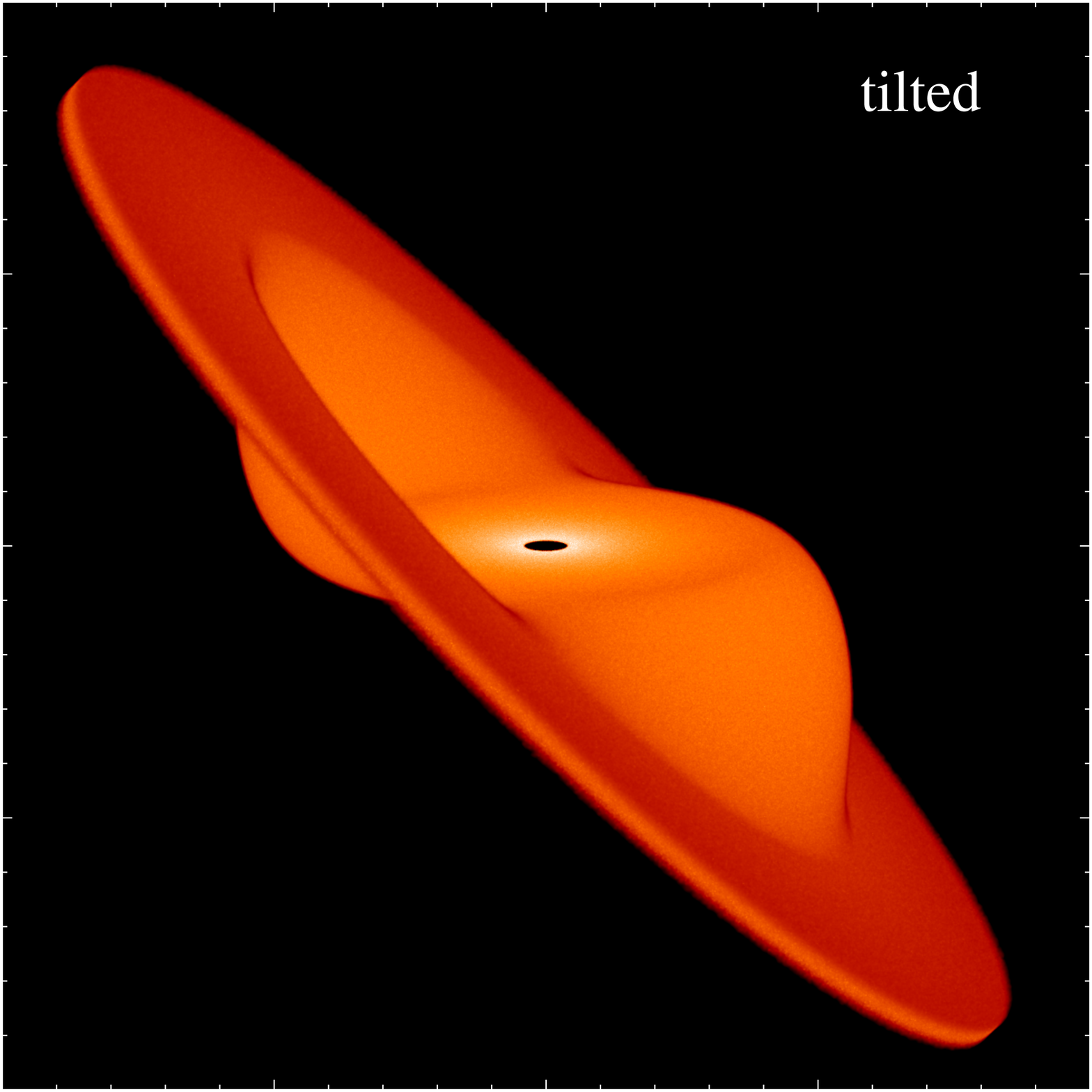}\hfill
  \includegraphics[width=0.47\textwidth]{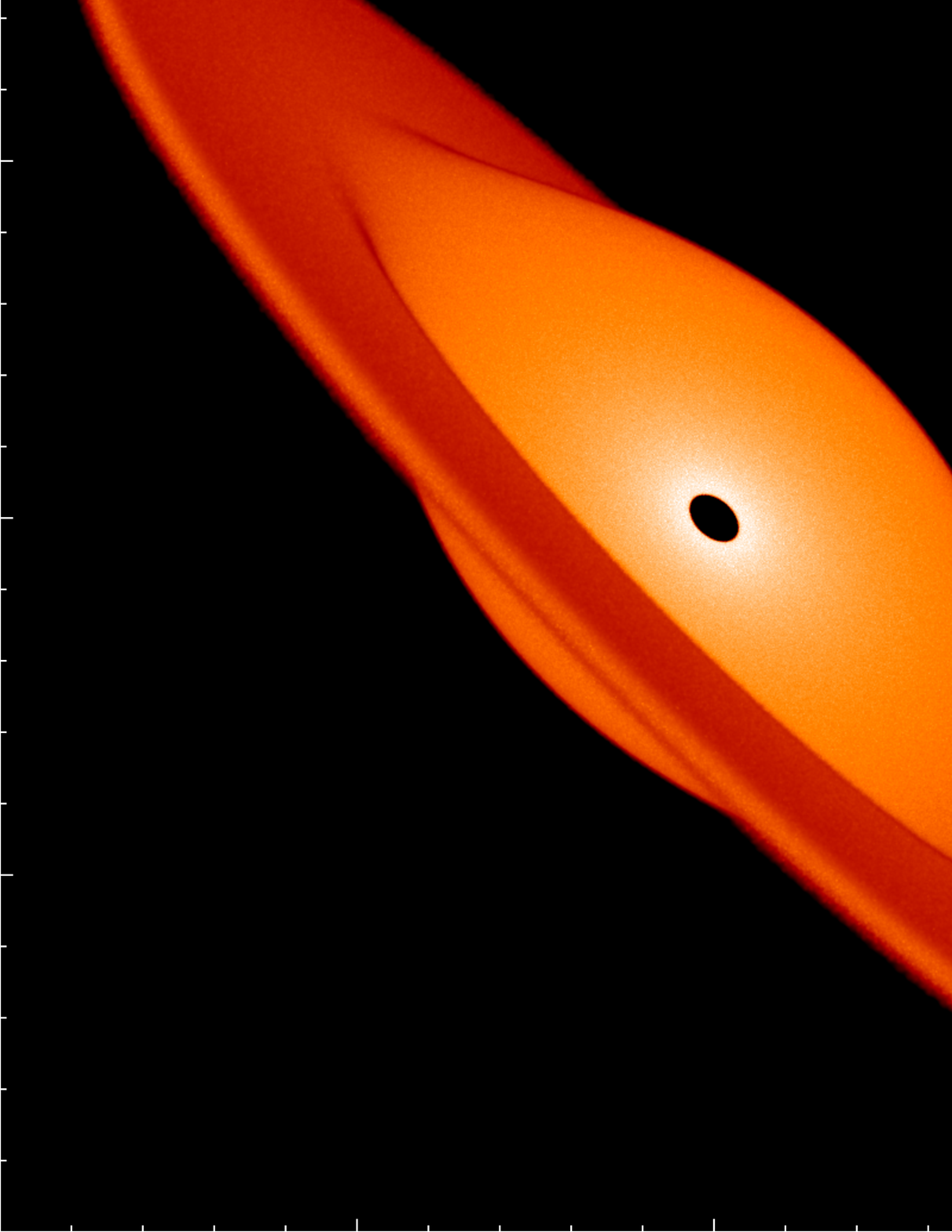}\vspace{0.1in}
  \hspace*{-0.05in}\begin{tikzpicture}
    \node (img1)  {\includegraphics[width=0.45\textwidth]{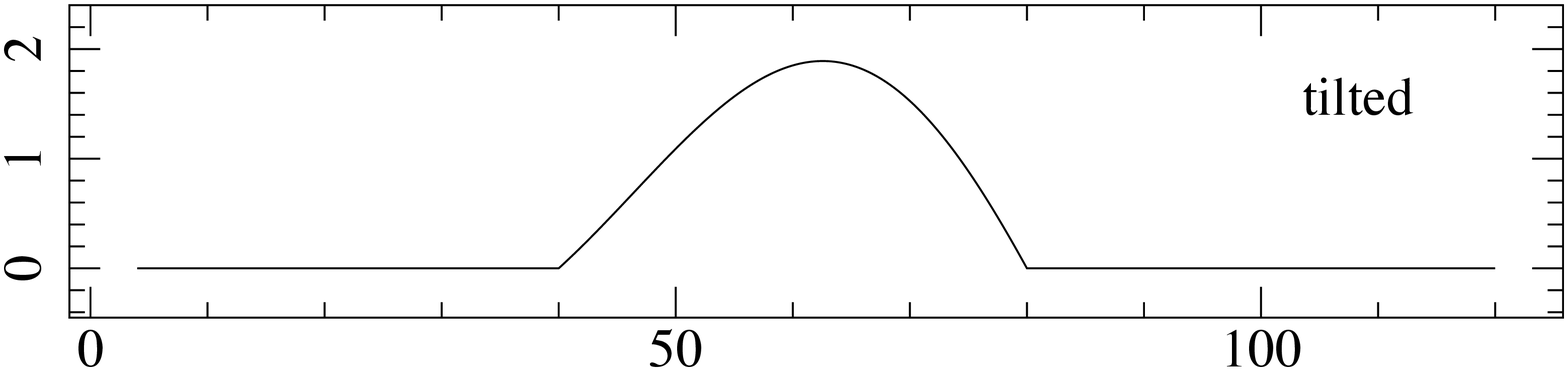}};
    \node[below of= img1, node distance=0cm, yshift=-1.2cm] {$R$};
    \node[left of= img1, node distance=0cm, rotate=90, anchor=center,yshift=4.3cm] {$\psi$};
  \end{tikzpicture}
  \hspace*{0.23in}
  \begin{tikzpicture}
    \node (img1)  {\includegraphics[width=0.45\textwidth]{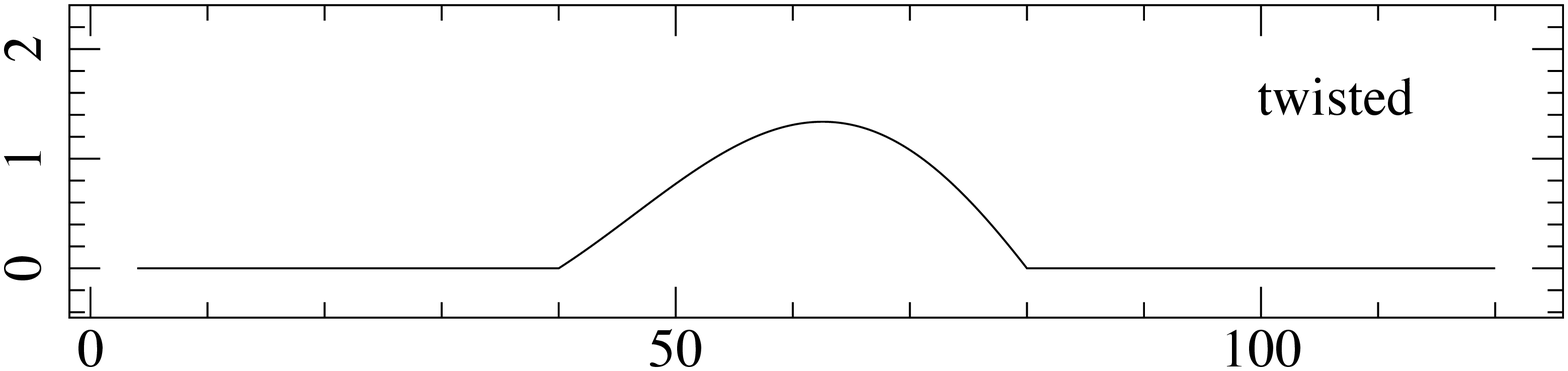}};
    \node[below of= img1, node distance=0cm, yshift=-1.2cm] {$R$};
    \node[left of= img1, node distance=0cm, rotate=90, anchor=center,yshift=4.3cm] {$\psi$};
  \end{tikzpicture}\vspace*{-0.2in}
  \caption{Example disc structures depicting the way in which a disc can be warped. On the left hand side we show a three dimensional rendering of a warp imposed on a disc where only the tilt angle, $\beta$, varies with radius. On the right hand side we show a similar structure, but this time only the disc twist angle, $\gamma$, is varied with radius. In each case the innner disc radius is at $R=4$, and the outer disc radius at $R=120$. For the tilted case, we take $\gamma = 0^\circ$ at all radii, while $\beta = 0^\circ$ for $R < 40$ and $\beta = 45^\circ$ for $R > 80$. Between these two radii we vary $\beta$ smoothly from $0^\circ$ to $45^\circ$ with a cosine bell. This results in the warp amplitude with radius, $\psi(R)$, shown in the plot below the image; in this case we have $\psi = R\left|\partial\beta/\partial R\right|$. For the twisted case, we take $\beta = 45^\circ$ at all radii, with $\gamma = 0^\circ$ for $R < 40$ and $\gamma = 45^\circ$ for $R > 80$. Between these two radii we vary $\gamma$ smoothly from $0^\circ$ to $45^\circ$ with a cosine bell. This results in the warp amplitude with radius, $\psi(R)$, shown in the plot below this image, and in this case we have $\psi = R\left|\sin\beta\,\partial\gamma/\partial R\right|$. We have shown these discs in an orientation in which the outer plane is the same in each case.}
  \label{Fig0}
\end{figure*}

The nomenclature for unstable warped discs, i.e.\@ `break' and `tear', is ill-defined in the literature. By a disc `break' we mean \citep[see also, for example,][]{Dogan:2018aa} that the disc has a sharp transition in either the disc tilt angle ($\beta$) or the disc twist angle ($\gamma$), and that this is usually accompanied by a sharp variation in the disc `surface' density\footnote{Note that for a planar disc orbiting in, say, the $x$-$y$ plane, the `surface' density is defined as the integral of the volume density with vertical height through the disc and is typically averaged in azimuth to give a single value for each radius. In a warped disc, the `surface' density retains this meaning, but now the `vertical' extent of the disc is taken normal to the local orbital plane and is a local quantity measuring the amount of mass orbiting per unit area at each radius. This is distinct from the column integral which would measure the amount of mass along any given sight line as seen by a distant observer.} at the same radius. Breaks may arise in a warped disc that achieves a sufficient warp amplitude and can occur with or without a forced precession of the disc orbits \citep[see, for example,][]{Nixon:2012aa}. The necessary conditions for disc `breaking' are discussed by \cite{Dogan:2018aa} and \cite{Dogan:2020aa}. By disc `tearing' we mean that an otherwise stable disc has been forced to precess (differentially) sufficiently rapidly that it achieves a large enough warp amplitude that renders the disc unstable to disc `breaking', and that the resulting parts of the broken disc are able to subsequently precess quasi-independently before interacting in a highly dynamic and variable fashion. With this nomenclature, we have that disc `breaking' is the underlying instability \citep[cf.][]{Dogan:2018aa}, and disc `tearing' refers to the action of breaking a disc with a forced precession and the subsequent dynamical behaviour of the unstable disc \citep[cf.][]{Nixon:2012ad}.

\cite{Larwood:1996aa} and \cite{Larwood:1997aa} present numerical simulations using smoothed particle hydrodynamics (SPH) of misaligned discs in binary systems, including both the circumbinary disc case and the case where the companion is external to the disc. In each case they show a simulation, typically the thinnest simulation presented, which exhibits disc tearing; the gravitational torque on the disc causes the disc orbits to precess and a break is formed between the inner and outer regions of the disc. They note that in these models the sound speed is too low to allow efficient communication between the different regions of the disc. Similar results are presented by \cite{Fragner:2010aa} who modelled the hydrodynamics using a grid code. These investigations exhibit disc tearing (breaking of the disc caused by a forced precession), but did not follow the evolution for long enough to explore the subsequent nonlinear, chaotic behaviour. The dynamical behaviour of disc tearing was explored by a series of papers in different contexts, namely; discs around spinning black holes \citep{Nixon:2012ad}, circumbinary discs \citep{Nixon:2013ab} and discs with an external binary companion \citep{Dogan:2015aa}. These simulations reveal the quasi-independent precession of rings of matter (some of which are radially narrow with $\Delta R \sim H$, and others more extended with $\Delta R \gg H$). These rings are found to precess until the internal angle between two neighbouring rings becomes large, at which point they can interact violently, with shocks between rings producing gas on low angular momentum orbits which can fall to smaller radii. This material can fall directly on to the central accretor or circularise at a new smaller radius. These processes can dramatically change the disc evolution and its observational character \citep[see, e.g.,][for discussion]{Nixon:2014aa}. Over the last few years disc tearing has been explored in the low viscosity case \citep{Nealon:2015aa}, has been found in general relativistic magnetohydrodynamic simulations \citep{Liska:2020aa}, and has been used to model observations of the circumstellar disc in GW Ori \citep{Kraus:2020aa}.

Recently, \cite{Dogan:2018aa} connected the disc breaking and tearing behaviour observed in numerical simulations with the instability of warped discs that was derived by \cite{Ogilvie:2000aa} through a local stability analysis of the warped disc equations. \cite{Ogilvie:1999aa} showed that in some circumstances as the warp amplitude is increased the local ``viscosity'' that resists the warping (e.g.\@ $\nu_2$; \citealt{Pringle:1992aa}) decreases. Thus the more warped the disc gets, the less it is able to resist warping further. This is found to occur for viscosities that are not much larger than $\alpha = 0.1$, and at a critical warp amplitude, $\psi_{\rm c}$, that depends on $\alpha$; typically lower $\alpha$ yields lower critical warp amplitudes. While \cite{Dogan:2018aa} focussed on discs with Keplerian rotation, \cite{Dogan:2020aa} explored the instability in the case of non-Keplerian rotation.

Here we perform numerical hydrodynamical simulations to explore the instability of warped discs and the disc tearing behaviour. To provide the forced precession we simulate discs around spinning black holes, and we vary the disc parameters including the viscosity parameter $\alpha$, the disc angular semi-thickness $H/R$ and the inclination of the disc $\theta$ with respect to the rotation axis of the black hole. In Section~\ref{numcon} we discuss the numerical methodology and expectations for the simulations. In Section~\ref{sims}, we report the results of the numerical simulations. In Section~\ref{conclusions} we present our conclusions.

\section{Numerical considerations}
\label{numcon}
In this section we discuss the methodology that is employed in our numerical simulations (reported in Section~\ref{sims} below), and how numerical effects may impact the results of our simulations. We use smoothed particle hydrodynamics (SPH) to model the accretion disc \citep[see e.g.][]{Price:2012aa}, and in particular we use the SPH code {\sc phantom} \citep{Price:2018aa}. To generate a time-dependent warp in the disc, with which we can analyse the evolution, we perform simulations of discs that are misaligned to a central spinning black hole. This includes both apsidal (Einstein) precession of eccentric disc orbits, and nodal (Lense-Thirring) precession of misaligned disc orbits.

\begin{figure*}
  \includegraphics[width=0.47\textwidth]{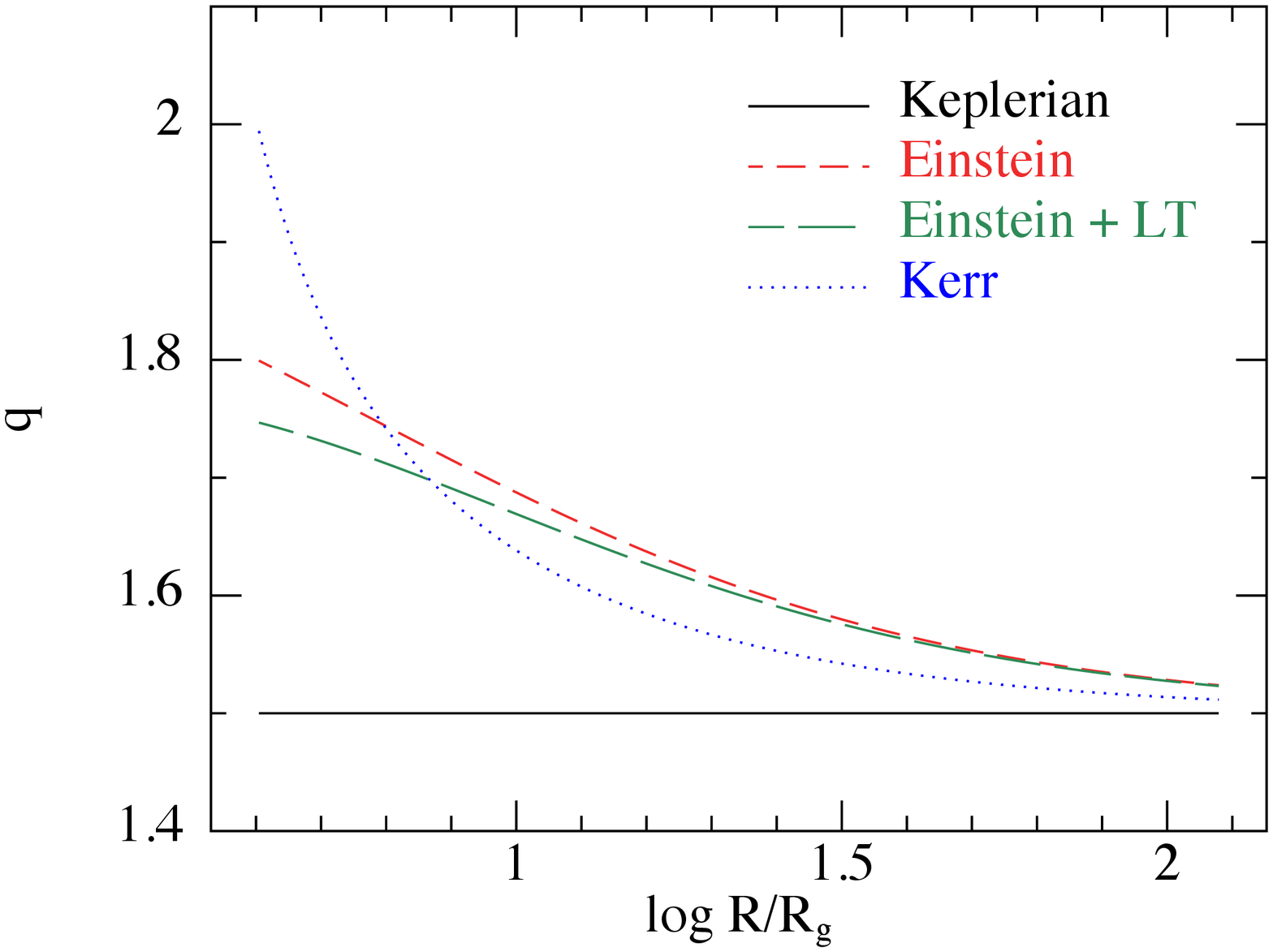}\hfill
  \includegraphics[width=0.47\textwidth]{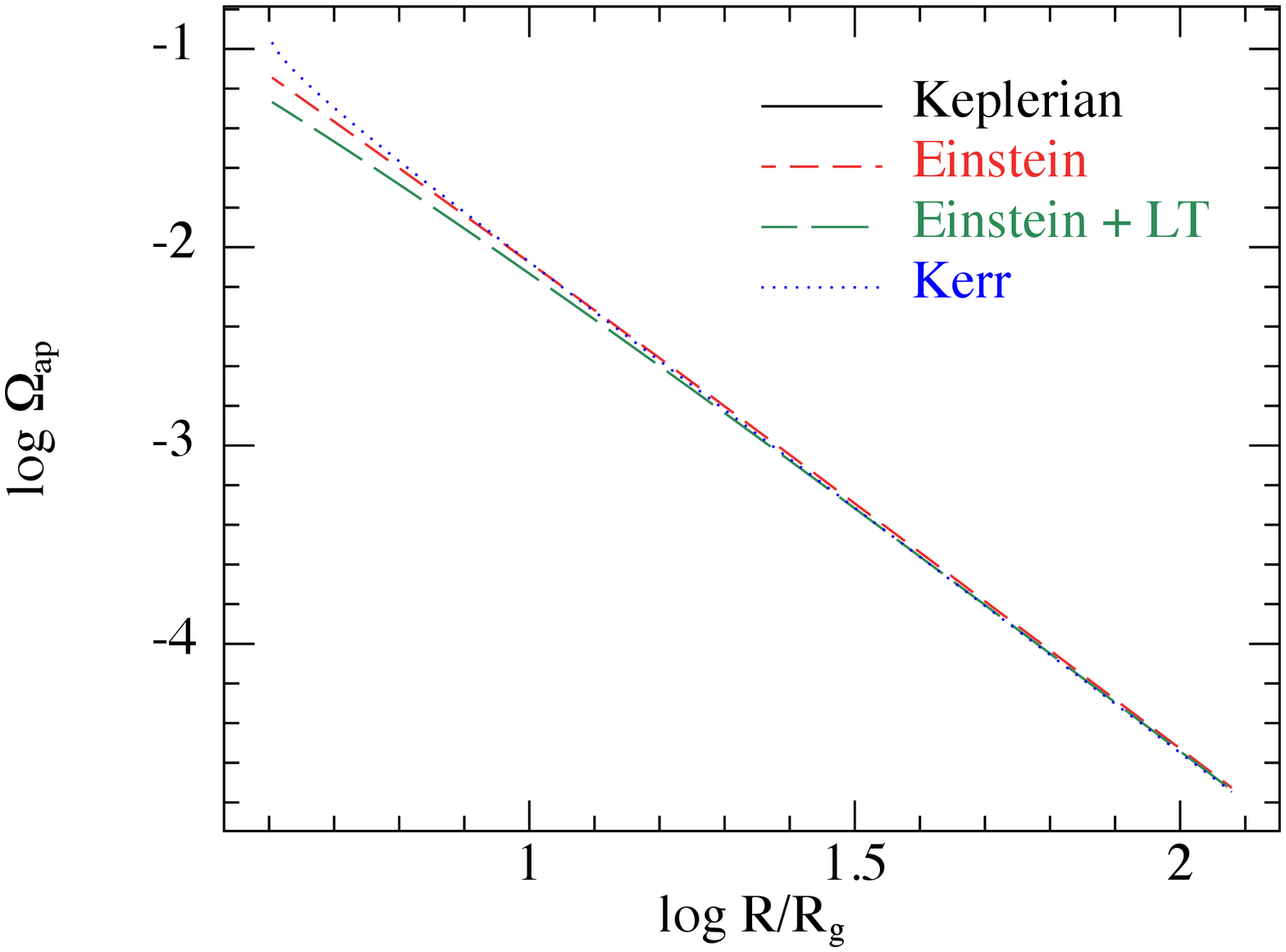}
  \caption{Orbital shear rate (left panel) and apsidal precession rate (right panel) for the Keplerian potential (black solid line), Einstein potential (red dashed line), Einstein potential with Lense-Thirring gravito-magnetic force (green long-dashed line) and from the Kerr metric (blue dotted line). The Keplerian shear rate is constant, with $q=1.5$, and the apsidal precession rate is zero in this case (not depicted). The curves have been evaluated for a black hole spin parameter of 0.5585. The equation for the Kerr shear parameter is taken from \cite{Gammie:2004aa}, while the apsidal precession rate is from \cite{Kato:1990aa}. The Einstein potential provides an accurate description of the apsidal precession rate. However, the shear parameter is only approximately correct with errors of the order of $10$ per cent.}
  \label{Fig1}
\end{figure*}

Following \cite{Nelson:2000aa} we employ a post-Newtonian approximation to model the gravity of the black hole. We use the Einstein potential for the apsidal precession with a gravito-magnetic force term to induce the required Lense-Thirring precession \citep[see][for details]{Nixon:2012ad,Price:2018aa}\footnote{We refer the reader to \cite{Liptai:2019aa} for a recent implementation of the SPH equations of motion in a General Relativity framework.}. While the Einstein potential provides a good description of the expected apsidal precession of the disc orbits, it does not produce an innermost stable circular orbit (ISCO) to define the inner edge of the disc. Instead we place an accretion radius at the location of the ISCO to truncate the disc there. The ISCO is defined by the location where circular orbits become unstable, which is equivalent to the square of the dimensionless epicyclic frequency, $\tilde\kappa^2 = \kappa^2/\Omega^2$, becoming negative. In terms of the orbital shear parameter, $q = -{\rm d\,ln}\Omega/{\rm d\,ln}R = (4-\tilde\kappa^2)/2$, this is where $q=2$. For the post-Newtonian approximation we employ, $q$ increases as the radius decreases, but $q < 2$ by the time the ISCO is reached. For the parameters we employ below, i.e. for a black hole spin of $a=0.5585$ for which $R_{\rm ISCO} = 4GM/c^2$, the shear parameter at the ISCO is $q\approx 1.75$. We plot the shear parameter and the apsidal precession rate for the Keplerian, Einstein and Kerr cases in Fig.~\ref{Fig1}. This shows that the Einstein potential accurately models the apsidal precession of disc orbits, but the local shear rate in the disc is only accurate to within approximately 10 per cent. The numerical method implemented in the {\sc phantom} SPH code to model Lense-Thirring precession has been shown to accurately reproduce the required precession rate (e.g. Fig. 17 of \citealt{Price:2018aa}). As shown by \cite{Dogan:2020aa} the growth rates of the instability depend on the value of the shear rate (see also Fig.~\ref{Fig2} below).

In numerical simulations it is important to distinguish between physical and numerical viscosity. By physical viscosity we refer to the \cite{Shakura:1973aa} viscosity that we employ here to model the angular momentum transport in discs that is generally thought to arise due to disc turbulence created by the magnetorotational instability \citep{Balbus:1991aa}. In different astrophysical systems the efficiency of angular momentum transport can differ, and this is usually encapsulated in the \cite{Shakura:1973aa} $\alpha$-parameter. In fully ionised discs $\alpha$ is typically found to be large \citep{King:2007aa,Martin:2019aa}, and may be lower in discs that are only partially ionised (see \citealt{Martin:2019aa} for a discussion). Here we employ several values of $\alpha$ to explore the possible range of disc evolution. This physical viscosity is implemented as a direct (Navier-Stokes) viscosity term in the simulations \citep[see Section 3.2.4 of ][and \citealt{Flebbe:1994aa}]{Lodato:2010aa}.

Numerical solution of the equations of hydrodynamics also leads to numerical viscosity (as is the case for any numerical method; see, for example, the discussion in the Appendix of \citealt{Sorathia:2013aa}). In the SPH method, numerical viscosity is added explicitly to the equations of motion as it is required to smooth discontinuities in the velocity field. We use the standard SPH artificial viscosity terms, with a linear term (with coefficient $\alpha_{\rm SPH}$) and a quadratic term (with coefficient $\beta_{\rm SPH}$). We take $\beta_{\rm SPH}=2$ and allow $\alpha_{\rm SPH}$ to vary following the switch proposed by \cite{Cullen:2010aa} with minimum value of 0.01 and maximum value of unity \citep[see][for details]{Price:2018aa}. These numerical terms contribute a small level of viscosity, which when computed in terms of a Shakura \& Sunyaev parameter is typically of the order of $\alpha^{\rm AV} \sim 0.01$, but note that this is resolution dependent (see equation~\ref{alphaAV} below). Ideally, this numerical viscosity should be small compared to the physical viscosity. Unfortunately this is not always the case, particularly when simulating low physical viscosity or thin discs. In these cases the numerical viscosity may play a role in determining the dynamics of the instability, especially when the critical warp amplitude and growth rates of the instability depend sensitively on the value of the disc viscosity (see, for example, Figure 3 of \citealt{Dogan:2018aa}).

It is also worth noting that in SPH simulations, the local resolution follows the local density of particles as the smoothing length (roughly the resolution lengthscale) $h \propto \rho^{-1/3}$, where $\rho$ is the density. This is a highly useful feature of the SPH method as it allows, for example, converging flows to be modelled with resolution that increases in regions of increasing density (e.g. gravitational fragmentation of a fluid). However, this feature also provides a challenge in modelling flows where the density drops significantly below the mean density. Such a drop in density occurs in warped discs in regions where the warp amplitude is high \citep{Nixon:2012aa}, and for an unstable disc which breaks into discrete rings the `surface density' is significantly reduced between the rings. Therefore it is possible that as the disc becomes unstable, the numerical viscosity could increase and become comparable to, or larger than, the physical viscosity in the simulation. However, as the instability is stabilised by larger viscosity \citep{Dogan:2018aa}, we expect this to have a stabilising effect on the simulations -- and thus if instability is observed, we expect that the disc is physically unstable\footnote{We note that this may not be true of simulations performed in the wavelike regime, where $\alpha < H/R$. In this case, the analytical work is not as well-established as in the diffusive case, and the stability criteria is not known. \cite*{Nealon:2015aa} report numerical simulations of discs with $\alpha < H/R$ and find that the disc can break. However, it is not clear that the numerical viscosity was sufficiently small that these simulations were in the wavelike regime at the time instability was found. We shall return to this question in the future.}. Such instability in low resolution simulations may not capture the correct growth rates or extent of the instability, but at high enough resolution (such that the physical viscosity is significantly larger than the numerical viscosity) it should be possible to capture the relevant dynamics.

In numerical simulations there will be a bulk viscosity present, which was not included in, for example, \cite{Dogan:2018aa} which employed $\alpha_{\rm b} = 0$. The bulk viscosity has an effect on the viscosity coefficients and thus on the stability and growth rates. \cite{Lodato:2010aa} report that the magnitude of the bulk viscosity present in an SPH simulation is $\alpha^{\rm AV}_{\rm b} \approx 5\alpha^{\rm AV}/3$, where $\alpha^{\rm AV}$ is the shear viscosity arising from the numerical viscosity\footnote{We note that an alternative method for modelling the physical viscosity in SPH simulations is to scale the numerical viscosity term to the appropriate value for a \cite{Shakura:1973aa} viscosity \citep[e.g.][]{Murray:1996aa,Lodato:2010aa}. In this method, the bulk viscosity will always be significant compared to the shear viscosity.}. The magnitude of the shear viscosity arising from the numerical viscosity in the continuum limit is \citep{Meru:2012aa}
\begin{equation}
  \label{alphaAV}
  \alpha^{\rm AV} = \frac{31}{525}\alpha_{\rm SPH}\frac{\left<h\right>}{H} + \frac{9}{70\pi}\beta_{\rm SPH}\left(\frac{\left<h\right>}{H}\right)^2\,,
\end{equation}
where $\left<h\right>/H$ is the shell-averaged smoothing length per disc scale-height. For a typical value of $\left<h\right>/H = 0.3$, and conservatively assuming $\left<\alpha_{\rm SPH}\right> \approx 1$ we have $\alpha^{\rm AV} \approx 0.02$, and thus $\alpha^{\rm AV}_{\rm b} \approx 0.03$.

In Fig.~\ref{Fig2} we provide the growth rates of the instability for $\alpha = 0.03$, $0.1$ and $0.3$ (the same as the values we employ in the simulations below) with comparison between $\alpha_{\rm b} = 0$ and $\alpha_{\rm b} = 0.03$. We also show the results for three different values of the disc shear, given by $q = 1.5$, $1.6$ and $1.7$. These plots show that the basic picture is unchanged; the discs are generally unstable at large warp amplitudes and small $\alpha$, while generally stable at small warp amplitudes and large $\alpha$ \citep[see][for details]{Ogilvie:2000aa,Dogan:2018aa,Dogan:2020aa}. However, the details are changed with the inclusion of non-zero bulk viscosity and non-Keplerian rotation. For example, the exact value of the critical warp amplitude can vary, and the growth rates at large warp amplitude can vary by factors of several.

\begin{figure*}
  \includegraphics[width=\textwidth]{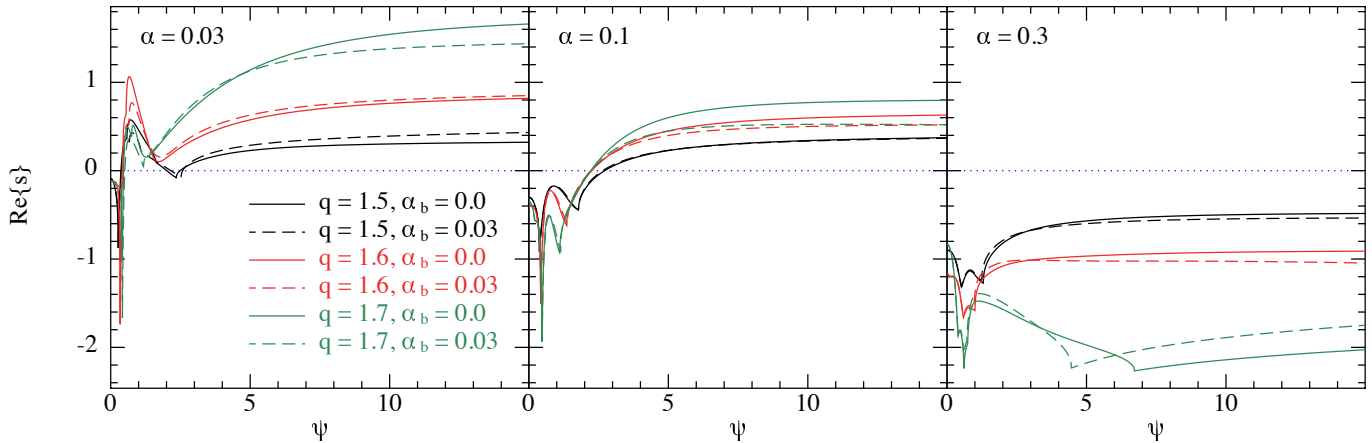}
  \caption{The dimensionless growth rates, $\Re\{s\}$, plotted against the warp amplitude, $\psi$, for different values of the viscosity parameter $\alpha$, the bulk viscosity $\alpha_{\rm b}$ and the shear parameter $q$. Recall that $s = -i(\omega/\Omega)(\Omega/c_{\rm s}k)^2$, and that perturbations grow ($\Re\{s\} > 0$) or decay ($\Re\{s\} < 0$) as $\exp(\Re\{-i\omega\}t)$. The values are calculated following the method outlined in \cite{Dogan:2018aa}, and using a code kindly provided by G.~I.~Ogilvie to calculate the torque coefficients \citep{Ogilvie:2013aa}. In the left, middle and right hand panels we plot the values for $\alpha=0.03$, $0.1$ and $0.3$ respectively. The black curves correspond to $q=1.5$ (Keplerian), the red curves to $q=1.6$ and the green curves to $q=1.7$. The solid curves represent zero bulk viscosity, while the dashed curves have $\alpha_{\rm b} = 0.03$. The blue dotted line represents zero growth rate, marking the line of instability; for positive (and real) growth rates the disc becomes unstable. For lower values of the disc viscosity, we see that the disc is unstable for a broad range of warp amplitude, and the growth rates increase with increasing $q$ (although, it is worth noting this trend does not increase to very large $q \approx 2$; \citealt{Dogan:2020aa}). For large disc viscosity, $\alpha=0.3$, we see that the disc is expected to be stable. A small level of bulk viscosity does not change the overall picture, but can affect the exact value of the critical warp amplitudes and growth rates.}
  \label{Fig2}
\end{figure*}

From the above it is clear that numerical modelling of the disc breaking instability is likely to yield subtle effects. In particular `convergence' of the simulations with increasing spatial resolution may not be easy to achieve. As we have noted above, the critical warp amplitudes and growth rates can depend sensitively on the value of the viscosity and thus may depend on the resolution. With modern day resolution the numerical viscosity in SPH simulations is not negligible, and for discs that are sufficiently thin ($H/R \ll 0.1$) to realistically model black hole accretion, we find typically the numerical viscosity is several 10s of per cent of the physical viscosity. For understanding the dynamics of these discs this is not a grave issue, as the net effect is that the simulations are somewhat more viscous than anticipated, i.e. if the numbers given above are representative, then for a physical $\alpha = 0.03$ we have a total simulated viscosity of $\alpha \approx 0.05$, while for $\alpha = 0.3$ we have $\alpha \approx 0.32$\footnote{It has been shown by \cite{Lodato:2010aa} that the standard SPH numerical viscosity behaves in an `isotropic' manner, that is to say it behaves like a Navier-Stokes viscosity, and thus one can add the viscosities in this way. This may not be true for other numerical methods.}. However, for interpreting the behaviour of simulations---for example, measuring the instability growth rates and comparing them with theoretical predictions---we must be aware that the numerical terms provide an additional, resolution-dependent effect. Therefore we anticipate that, for simulations in which the disc is `resolved' (that is $\left<h\right>/H < 1$) as the resolution of the simulations is increased we should recover the same basic dynamics (such as whether the disc is stable or unstable and whether the disc aligns to the black hole spin or not), but the precise details (such as the exact location of disc breaks and the exact growth rate in the unstable region) may vary\footnote{It is also worth noting that the disc has $\rho \rightarrow 0$ at the inner boundary. This region is therefore difficult to resolve in an SPH simulation, and is where the Lense-Thirring torque is most strongly applied. However, we remark that while we have discussed the drawbacks of the numerical method we employ here, it is worth bearing in mind that any numerical method suffers drawbacks. For example, grid based methods struggle to conserve angular momentum, particularly for orbits which cross the symmetry plane of the grid which is inevitable for a warp.}.
 
With these caveats in mind we return to the focus of the paper, namely to compare numerical simulations of warped discs with the instability of warped discs found by analytical calculations to occur at low viscosity and large warp amplitude \citep{Ogilvie:2000aa,Dogan:2018aa}. We note here, that the analytical work on this instability takes the form of a local linear stability analysis which makes use of the assumption that the perturbations grow (or decay) on timescales short compared with the evolution of the unperturbed state. In a dynamic simulation in which the background state is continuously evolving, it is likely that the instability is only realised in situations where the growth rate is sufficiently high that the instability has time to act before the disc evolves further. In this sense, the solutions presented by \cite{Dogan:2018aa} represent necessary, but not sufficient, criteria for instability, with the sufficiency supplied by the condition that the background state evolve more slowly than the growth of unstable modes. Taking the above into account our aims for the simulations presented in this paper are to show, in agreement with the analytical predictions of the warped disc instability, that
\begin{enumerate}
\item discs with warp amplitude above the critical warp amplitude, $\psi > \psi_{\rm c}$, for extended periods of time lead to instability,
\item discs with $\psi < \psi_{\rm c}$ are stable, and
\item that the growth rate of the warp amplitude in the unstable regions follow the general trends of the predicted growth rates, i.e. that the growth rates are generally higher for smaller viscosity and depend on the warp amplitude.
\end{enumerate}

\section{Numerical simulations}
\label{sims}
We present three dimensional hydrodynamical simulations of accretion discs with an imposed external torque. The fluid dynamics is modelled with the smoothed particle hydrodynamics technique \citep[SPH; e.g.\@][]{Price:2012aa} using the publicly available code {\sc phantom} \citep{Price:2018aa}. This code was first used to model warped discs by \cite{Lodato:2010aa}, who simulated isolated warped discs with no external torque and found excellent agreement between the disc evolution and the evolution predicted by the analytical theory of \cite{Ogilvie:1999aa}. Motivated by the broken discs found in one dimensional calculations of the disc structure by \cite{Nixon:2012aa}, {\sc phantom} was subsequently used to model the behaviour of broken discs by \cite{Nixon:2012ab}. Discs with an external torque were modelled in the Lense-Thirring case \citep{Nixon:2012ad}, the circumbinary case \citep[][see also \citealt{Facchini:2013aa}]{Nixon:2012ac,Nixon:2013ab}, and the circumstellar case with a binary companion \citep{Dogan:2015aa}. In each case the dynamics proceeds in a similar manner; modest inclinations lead to a mild warp in the disc and alignment with the perturbing torque, while large inclinations generally lead to strong warps and often the disc breaks into discrete planes which can subsequently precess independently -- this process was named as disc tearing by \cite{Nixon:2012ad}.

Here, we return to the case of discs that are misaligned to the rotation of a spinning black hole, and thus precess due to the Lense-Thirring effect. As discussed above, we model the central potential with the Einstein potential \citep[e.g.\@][]{Nelson:2000aa} to take account of the apsidal precession of disc orbits, and we include the Lense-Thirring term through a gravito-magnetic force term \citep[e.g.\@][]{Nelson:2000aa}. For details of the implementation see \cite{Nealon:2015aa} and \cite{Price:2018aa}.

We take the black hole spin to be $a = \frac{2}{3}(4-\sqrt{10}) \approx 0.5585$  \citep[see][]{Lubow:2002aa} which yields an ISCO at $4R_{\rm g}$, where $R_{\rm g} = GM/c^2$. The inner edge of the disc, $R_{\rm in}$, is set to the ISCO radius, and the outer edge of the disc is initially at $120R_{\rm g}$. The initial disc surface density follows a power-law that accounts for the zero-torque inner boundary condition as matter accretes on to the black hole, given by
\begin{equation}
  \label{sigr}
  \Sigma(R) \propto (R/R_{\rm in})^{-p}\left(1-\sqrt{R_{\rm in}/R}\right)\,,
\end{equation}
where the normalisation is set by the total disc mass\footnote{As we do not take account of the gas self-gravity in these simulations, or allow the back reaction on the black hole spin vector (from accretion or the Lense-Thirring torque), the exact value of the disc mass plays no role in the calculation.}. The disc sound speed follows the locally isothermal approximation with
\begin{equation}
  c_{\rm s}(R) \propto (R/R_{\rm in})^{-q_{\rm s}}\,,
\end{equation}
with the normalisation fixed by the value of the disc angular semi-thickness, $H/R$, which is specified at $R_{\rm in}$. For the simulations presented here, we take the power-law indices to be $p=1$ and $q_{\rm s}=0.5$, with the latter giving a constant disc angular semi-thickness. As discussed above there are two components to the viscosity in the simulated disc, a numerical component and a physical component. For the numerical viscosity we use the switch proposed by \cite{Cullen:2010aa} with the linear numerical viscosity coefficient $\alpha_{\rm SPH}$ allowed to vary between 0.01 and unity, and we employ a quadratic numerical viscosity with $\beta_{\rm SPH}=2$. For the physical viscosity, we impose a Navier-Stokes (isotropic) viscosity of magnitude given by the Shakura-Sunyaev shear viscosity $\alpha$ \citep[see Section 3.2.4 of][for details]{Lodato:2010aa}. Unless stated otherwise, the simulations presented below employed $N_{\rm p} = 10^7$ particles. The simulations are evolved to a time of $10^5 GM/c^3$, which corresponds to a factor of a few less than the Lense-Thirring (nodal) precession timescale at the outer disc radius (to ensure the outer regions are undisturbed on the timescale of the simulation) and is long enough to ensure the inner disc regions have evolved significantly (the run time is equivalent to a Lense-Thirring precession timescale at a radius of approximately $50R_{\rm g}$). 

In the following sections we present quantities from the simulations that require averaging over the SPH particle data. For example, the warp amplitude at each radius requires taking the radial derivative of the unit tilt (orbital angular momentum) vector for the disc. To calculate the angular momentum vector as a function of radius in the disc, we must take an average over a selection of the disc particles. For the plots presented below, we split the disc into $N$ shells spaced logarithmically in radius. For each shell, with spherical radius $r_{\rm s}$, we average over the particles that have $|r-r_{\rm s}| < \xi H$, where $H$ is the disc scale-height at $r_{\rm s}$. We find that choosing $N=150$ and $\xi=2$ serves to minimise Poisson noise from the discreteness of the particles, while not oversmoothing features in the disc.

We investigate the warped disc dynamics by varying the following parameters; we simulate discs with two thicknesses ($H/R = 0.02, 0.05$), three inclinations ($\theta = 10^\circ, 30^\circ$ \& $60^\circ$), and three values of the disc viscosity ($\alpha = 0.03, 0.1$ \& $0.3$). This is a total of eighteen simulations, so rather than provide eighteen detailed sets of results we instead focus our attention on presenting the important results. In the following sub-sections we provide the results of varying each of these parameters in turn, and then also more detailed investigation into the properties of the simulations that exhibit disc tearing. We note that in all of the simulations we find that when the disc is stable the solutions vary slowly with time (perhaps with some propagating waves in the cases of lowest viscosity), while for the unstable discs the solutions are time-dependent; in this case at times the discs are broken and this can be long-lived, short-lived or repeating behaviour. We discuss this behavior in detail below.

\subsection{Varying the disc thickness}
\label{varyhonr}
\begin{figure*}
  \includegraphics[width=0.33\textwidth]{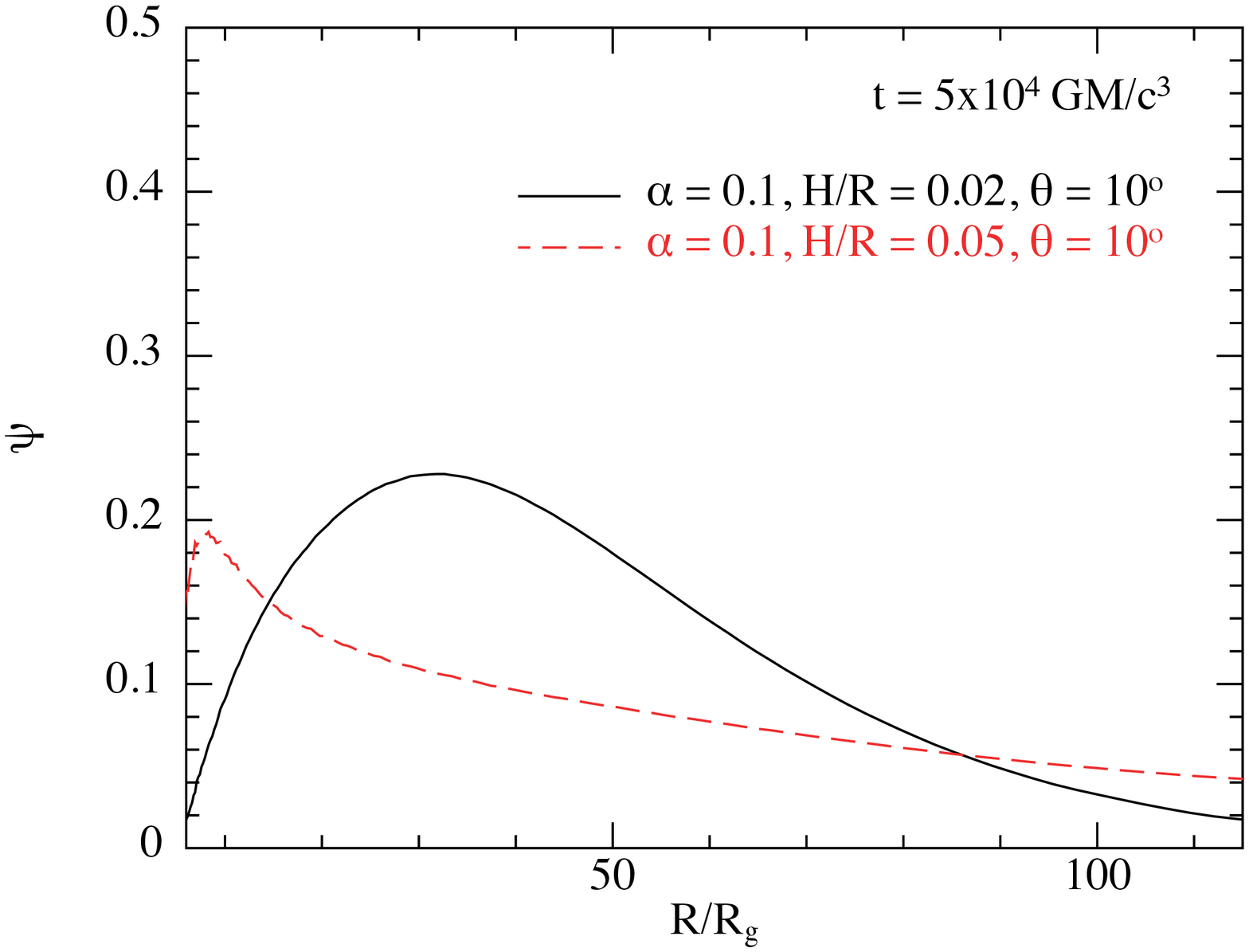}\hfill
  \includegraphics[width=0.33\textwidth]{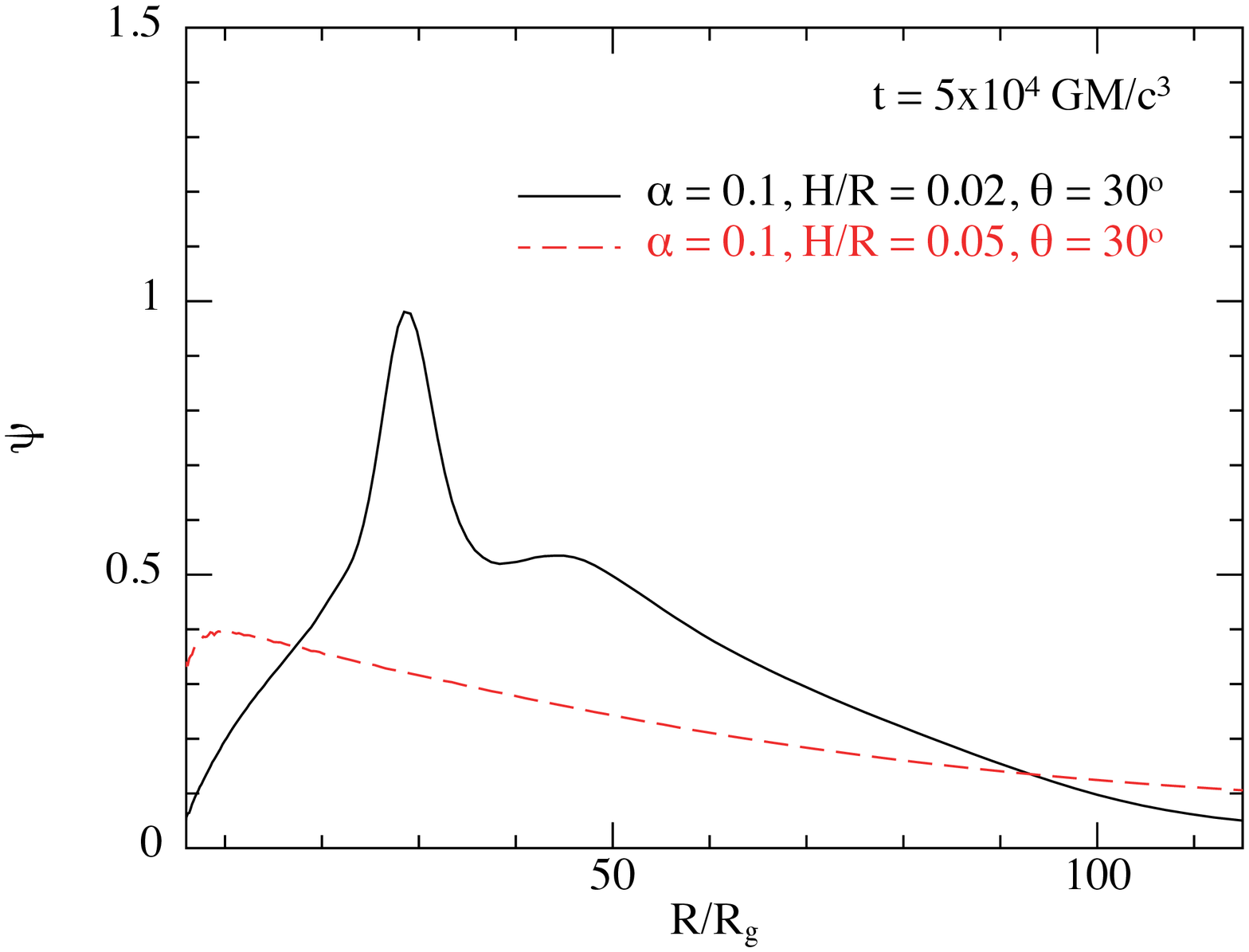}\hfill
  \includegraphics[width=0.33\textwidth]{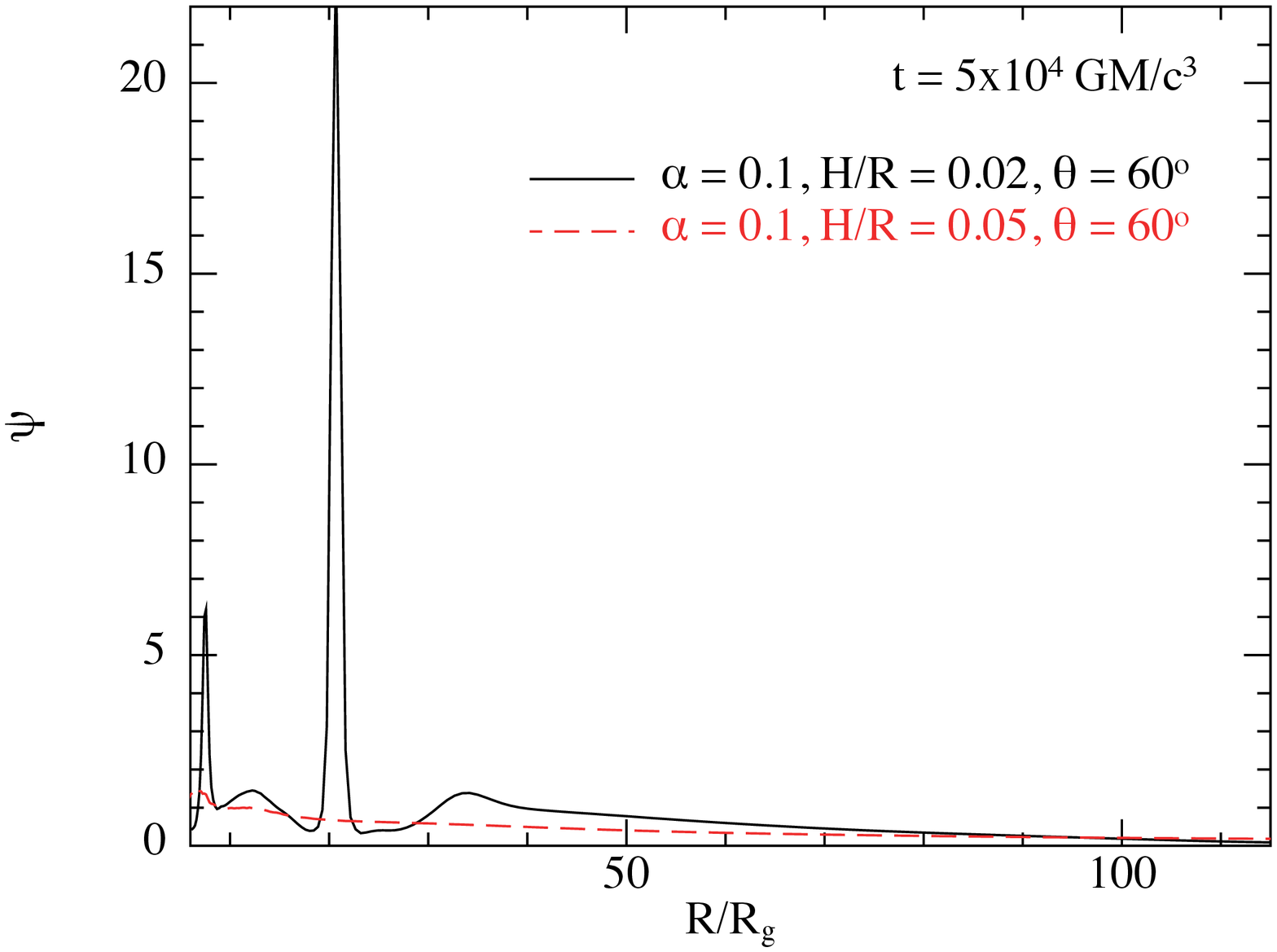}
  \caption{The warp amplitude, $\psi = R\left|\partial\mathbi{l}/\partial R\right|$, plotted against radius, $R/R_{\rm g}$, at a time of $5\times10^4 GM/c^3$ for the simulations with $\alpha=0.1$. Each panel compares the warp amplitude in the simulation with $H/R=0.02$ (black solid line) with the corresponding simulation with $H/R=0.05$ (red dashed line), with the left(middle)[right] hand panel showing the simulation with the disc initially inclined by $10^\circ$($30^\circ$)[$60^\circ$]. In each panel the range of warp amplitude that is plotted is varied to show the full extent of the solution. In each case the peak of the warp amplitude moves to smaller radius as the disc thickness is increased, but the peak warp amplitude is typically higher when the disc is thinner. For the simulation with $H/R = 0.02$ and $\theta = 60^\circ$, there are two clear breaks in the disc at $R/R_{\rm g} \approx 8$ and $R/R_{\rm g}\approx 20$, and these are typically accompanied by deficits of warp amplitude in neighbouring regions. This figure also shows that the critical warp amplitude at which the disc breaks is larger than unity for $\alpha = 0.1$ as the disc remains stable for $\theta = 30^\circ$. For $\theta = 60^\circ$, the disc typically becomes unstable when $\psi \gtrsim 2$, consistent with $\psi_{\rm c} \approx 2.25$ in this case.}
  \label{Fig3}
\end{figure*}

The disc thickness is expected to affect the evolution of warped discs. For a disc subject to an external torque, the disc thickness affects the magnitude of the effective viscosities in the disc (which are proportional to $(H/R)^2$) and thus affects the location at which the external torque becomes dynamically important. For example, the disc tearing radius derived by \cite{Nixon:2012ad} depends on the disc thickness, with a larger extent of the disc expected to be vulnerable to instability if the disc is thinner. Another effect that arises from the disc thickness is that it alters the lengthscale on which the disc can be warped. It is not possible to bend a disc on a lengthscale shorter than $\sim H$, and thus when the same torque is applied thick discs cannot achieve as large a warp amplitude as thin discs. This implies that thick discs are less vulnerable to instability as they are less likely to achieve a warp amplitude that is greater than the critical warp amplitude \citep{Dogan:2018aa}. We demonstrate this by plotting in Fig.~\ref{Fig3} the warp amplitude with radius for the simulations with $\alpha = 0.1$ and the comparison between $H/R = 0.02$ and $H/R = 0.05$ (with the left hand panel at $\theta = 10^\circ$, middle panel at $30^\circ$ and the right hand panel at $\theta = 60^\circ$). In these cases the discs are all coherently warped, except for the thinnest and most inclined case ($H/R=0.02$, $\theta=60^\circ$) which exhibits two breaks at $R/R_{\rm g} \approx 8$ and $R/R_{\rm g}\approx 20$. We note that the plots are made after a time of $5\times10^4 GM/c^3$, which is halfway through the simulation. This corresponds to a Lense-Thirring precession timescale at $R \approx 35R_{\rm g}$. The main results depicted by these plots is that (1) thicker discs have a peak in the warp amplitude that occurs at a smaller radius than thinner discs \citep[see also, for example,][]{Kumar:1985aa,Natarajan:1998aa}\footnote{It is worth noting that the fact that the warp amplitude can peak near the inner edge of the disc is of interest, as it allows for the possibility that the disc can produce time-dependent, repeating, behaviour there. If the warp amplitude is maximal near the inner edge, and the value of the warp amplitude is above the critical warp amplitude there \cite[cf.][]{Dogan:2020aa}, then the innermost ring of the disc may tear off, precess and be rapidly accreted, and this process may repeat indefinitely (or at least until the disc conditions change, causing the warp amplitude there to evolve). We discuss one example of this from our simulations in more detail in \cite{Raj:2021ab}.}, and that (2) in general (but not always\footnote{For example, the simulations corresponding to the left hand panel of Fig.~\ref{Fig3} but with $\alpha = 0.03$ (not depicted) show a higher peak warp amplitude in the thicker case. For these cases, the peak in the thicker case occurs at a small radius, at which the thinner disc has already aligned to the black hole spin.}) the peak warp amplitude is higher for thinner discs and this occurs across a broad section of the disc. For the right hand panel of Fig.~\ref{Fig3} we can see that for the same viscosity parameter and disc inclination, the disc thickness can play a key role in determining how vulnerable the disc is to instability; this can occur both due to the increased warp amplitude, and also due to the increased local viscous timescale, associated with thinner discs. We also note that for these parameters, the thinner disc ($H/R=0.02$) is aligned to the black hole spin plane at the ISCO, but that this is not true for the thicker case ($H/R=0.05$) where, while there is a downturn in the inclination, typically the disc is still significantly misaligned at the ISCO.

\subsection{Varying the disc inclination}
\label{varytheta}
\begin{figure*}
  \centering
  \includegraphics[width=0.40\textwidth]{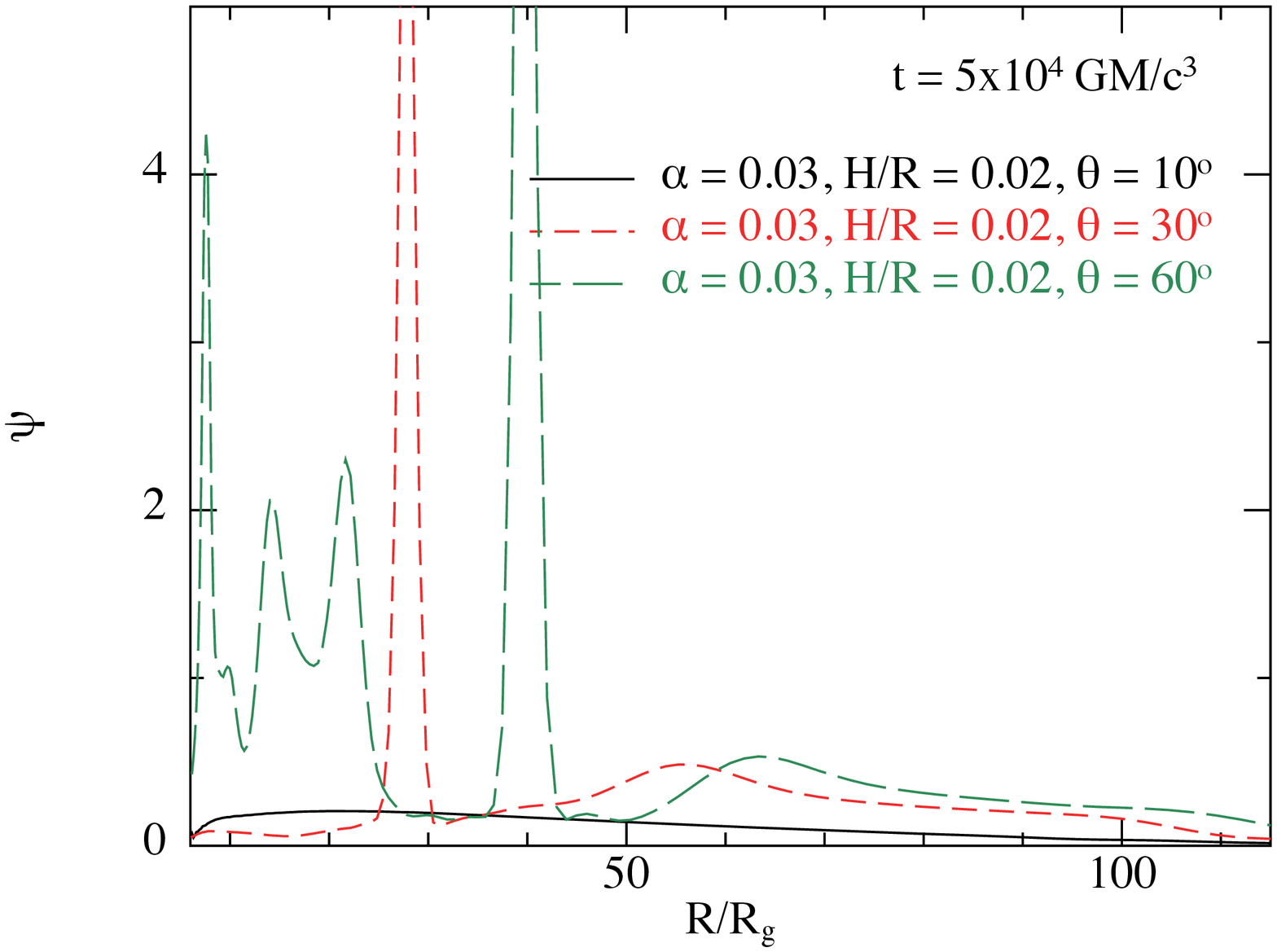}\hspace{1in}
  \includegraphics[width=0.40\textwidth]{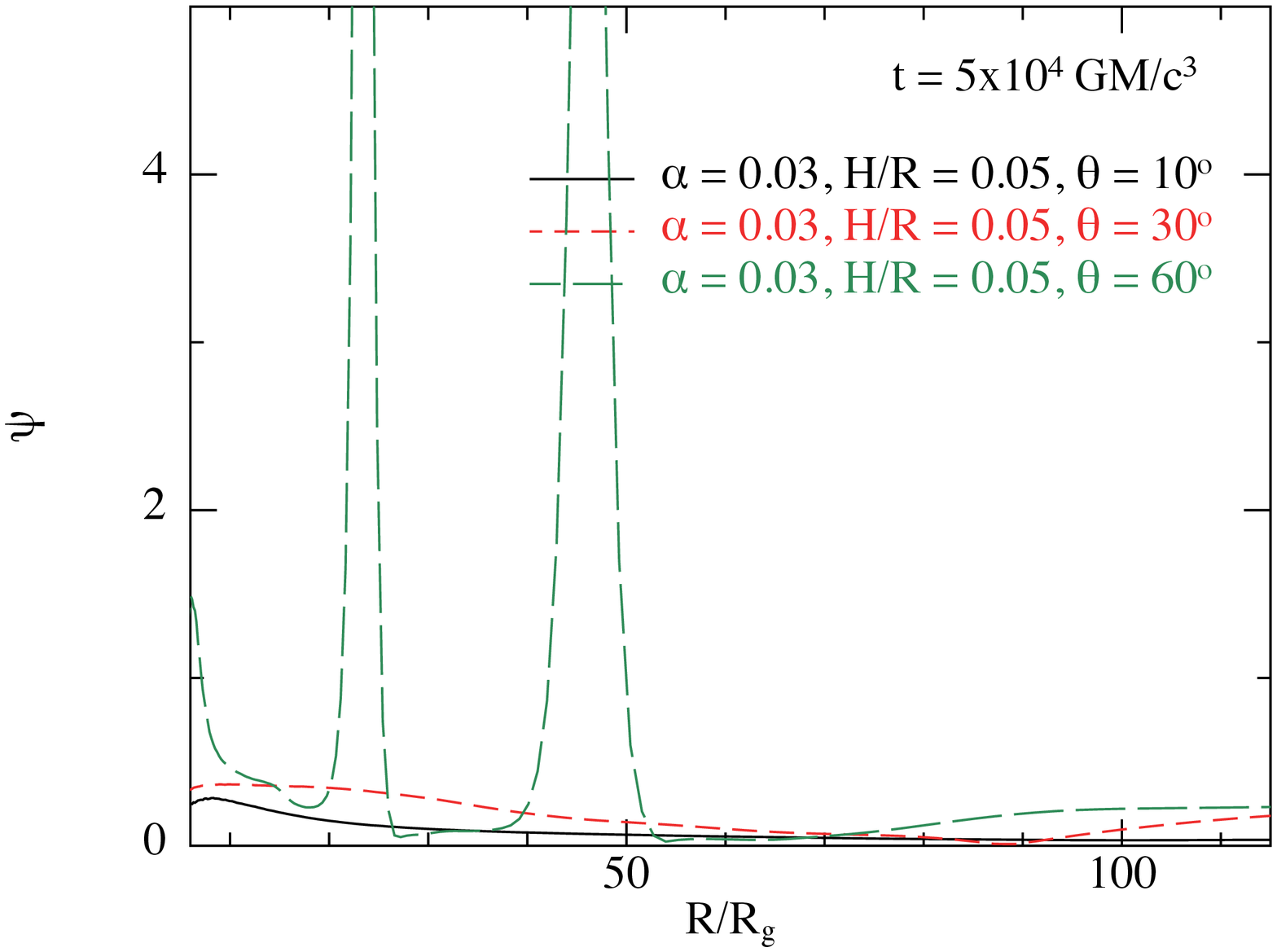}
  \caption{The warp amplitude, $\psi = R\left|\partial\mathbi{l}/\partial R\right|$, plotted against radius, $R/R_{\rm g}$, at a time of $5\times10^4 GM/c^3$ for the simulations with $\alpha=0.03$. Each panel compares the warp amplitude in the simulation with an inclination of $\theta=10^\circ$ with the corresponding simulation with $\theta=30^\circ$ and $\theta=60^\circ$, with the left(right) hand panel showing the simulations with disc thickness of $H/R = 0.02(0.05)$. The warp amplitude is typically larger for larger inclinations (and also for thinner discs; cf. Fig.~\ref{Fig3}), and thus these discs are more likely to exhibit disc tearing. We note that for $\alpha = 0.03$ the critical warp amplitude is $\psi_{\rm c} \approx 0.5$.}
  \label{Fig4}
\end{figure*}

The disc inclination has an effect on the disc evolution as, for larger inclinations, the warp amplitude is typically higher. This is most easily seen for low inclinations where the evolution is essentially linear, in that the behaviour can be rescaled by the inclination angle $\theta$ \citep[cf.][]{Lubow:2002aa}. However, once the inclination angle becomes large enough (typically greater than a few times $H/R$, which for our parameters is $\gtrsim 10^\circ$) then the evolution becomes noticeably non-linear in the inclination angle. This is evident in Fig.~\ref{Fig4} where the disc behaviour shows a strong dependence on inclination angles between $\theta = 10^\circ$, $30^\circ$ and $60^\circ$. This figure shows the warp amplitudes, with the same format as Fig.~\ref{Fig3}, but this time for the simulations with $\alpha = 0.03$; the left hand panel shows $H/R = 0.02$ while the right hand panel shows $H/R=0.05$ and in both cases the black-solid(red-dashed)[green-long-dashed] line corresponds to $\theta = 10^\circ(30^\circ)[60^\circ]$. For the lowest inclination of $10^\circ$ the warp amplitude remains low, and the disc attains a coherent warped shape. This is also true for the $30^\circ$ case when $H/R=0.05$. For the other cases, the disc becomes sufficiently warped that the warp amplitude rises above the critical warp amplitude \citep{Dogan:2018aa} and the disc tears into discrete rings (the rapid change in radius of the disc orbital angular momentum vector is indicated in the warp amplitude by the sharp peaks present in Figs~\ref{Fig3} \& \ref{Fig4}). These results reinforce the conclusion from the previous section (\ref{varyhonr}) as the thinner discs are subject to more vigorous tearing, and also confirm the expectation that larger inclination results in higher warp amplitude and a higher susceptibility to instability. We note that to first order the Lense-Thirring precession that we simulate here has a frequency that is independent of the inclination angle (i.e. the torque is proportional to $\sin\theta$), while the precession frequency induced by, for example, the gravitational field of a misaligned binary system \citep[see, e.g.,][]{Bate:2000aa} has an angular dependence such that the frequency goes to zero at $\theta=90^\circ$ (the torque in this case is proportional to $\sin2\theta$). This means that for discs in binary systems the increase in warp amplitude with increasing inclination angle peaks at $45^\circ$ (and also $135^\circ$), rather than at $90^\circ$ in the Lense-Thirring case.

\subsection{Varying the disc viscosity}
The disc viscosity parameter $\alpha$ plays a crucial role in the dynamics of warped discs. For $\alpha < H/R$ the disc response to warping is to propagate bending waves, while for $\alpha > H/R$ the disc responds by diffusing the warp \citep{Papaloizou:1983aa}. In the diffusive case, which we focus on here (although note that some of our simulations are potentially wavelike with $\alpha$ slightly smaller than $H/R$), the exact value of $\alpha$ plays two distinct roles in the dynamics.

The first is that the torque coefficients for modest warp amplitudes vary with $\alpha$. In the limit of small warps and low $\alpha$, the usual `planar' viscosity is proportional to $\alpha$, while the `vertical' viscosity is proportional to $1/\alpha$ \citep{Papaloizou:1983aa}\footnote{Here we have assumed that the underlying mechanism generating the viscosity is essentially isotropic such that any shear is damped at the same rate (given by $\alpha$), such as might be expected when the viscosity arises due to small scale turbulence. For a discussion of the viscosity coefficients in the case that the underlying mechanism generates an anisotropic viscosity, see \cite{Nixon:2015ab}.}. Thus for small values of $\alpha$, there is a significant discrepancy between the timescales associated with the disc resisting a warp by attempting to straighten itself (governed by the vertical viscosity) and the timescale on which mass flows inwards through the disc (governed principally by the planar viscosity). However, when $\alpha$ is large and approaches unity, these timescales become comparable. This means that large $\alpha$ makes it difficult to break a disc; this is because in this case the timescale on which the disc can internally rearrange itself, and potentially locally flatten the disc into two (or more) distinct planar regions, is similar to the timescale on which matter flows from one region of the disc to another; this latter effect can keep the disc as a coherent whole. Whereas for smaller $\alpha$ the radial communication in the disc can be severely restricted at large warp amplitudes, allowing discrete parts of the disc to behave independently \cite[see][for discussion]{Nixon:2012aa}.

The second role that the disc viscosity plays in warped disc dynamics is more subtle. \cite{Ogilvie:1999aa} showed that the torque coefficients vary as a function of the warp amplitude in the disc, and that the way in which they vary also depends on $\alpha$. \cite{Ogilvie:2000aa} showed that the variation in torque coefficients with warp amplitude causes an instability, and \cite{Dogan:2018aa} and \cite{Dogan:2020aa} explored this instability and connected it with the disc breaking/tearing that had been found in numerical simulations. These analyses show that the instability is generally more vigorous (lower critical warp amplitudes and higher growth rates) for lower values of $\alpha$.

\begin{figure*}
  \centering
  \includegraphics[width=0.40\textwidth]{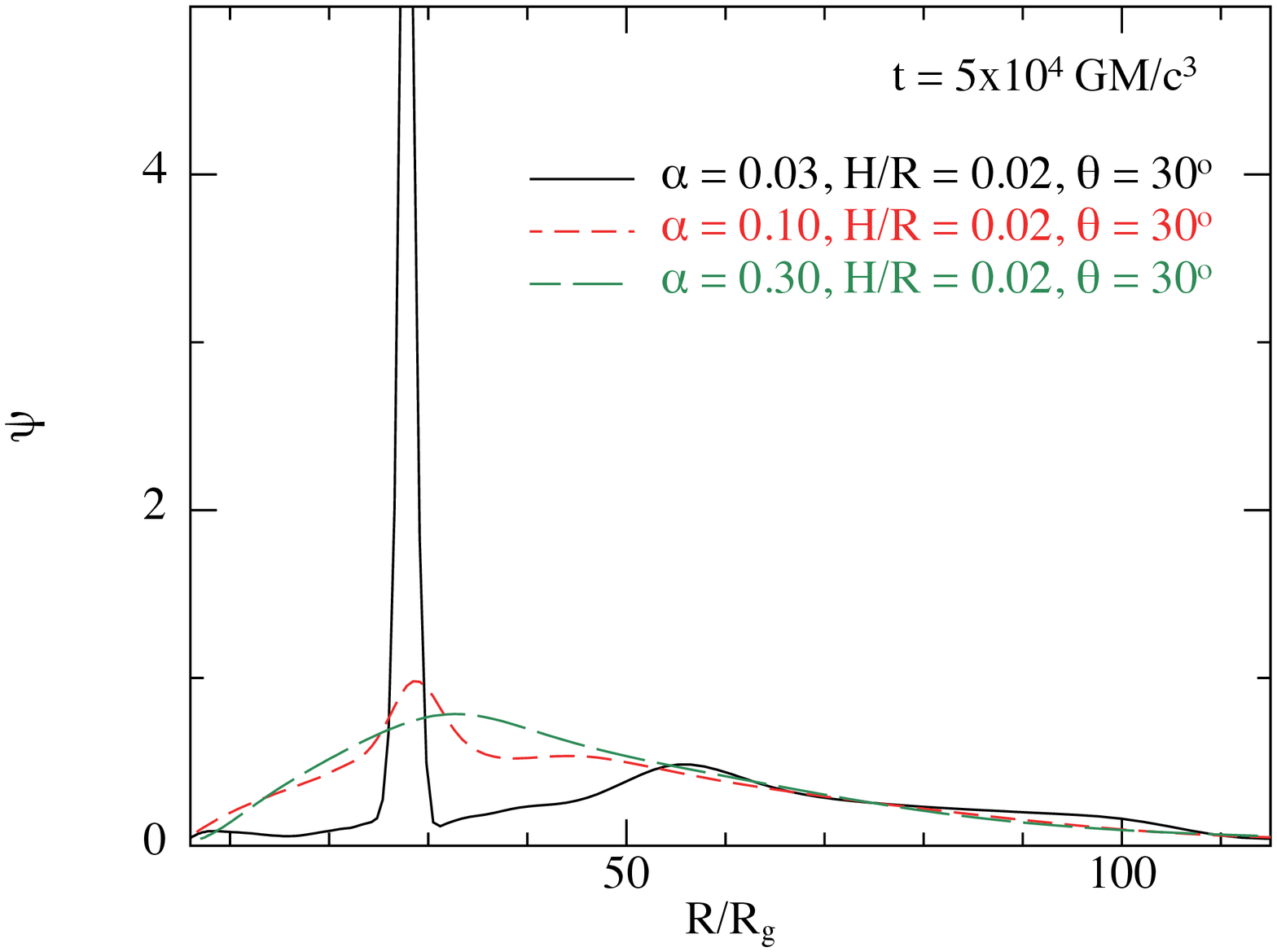}\hspace{1in}
  \includegraphics[width=0.40\textwidth]{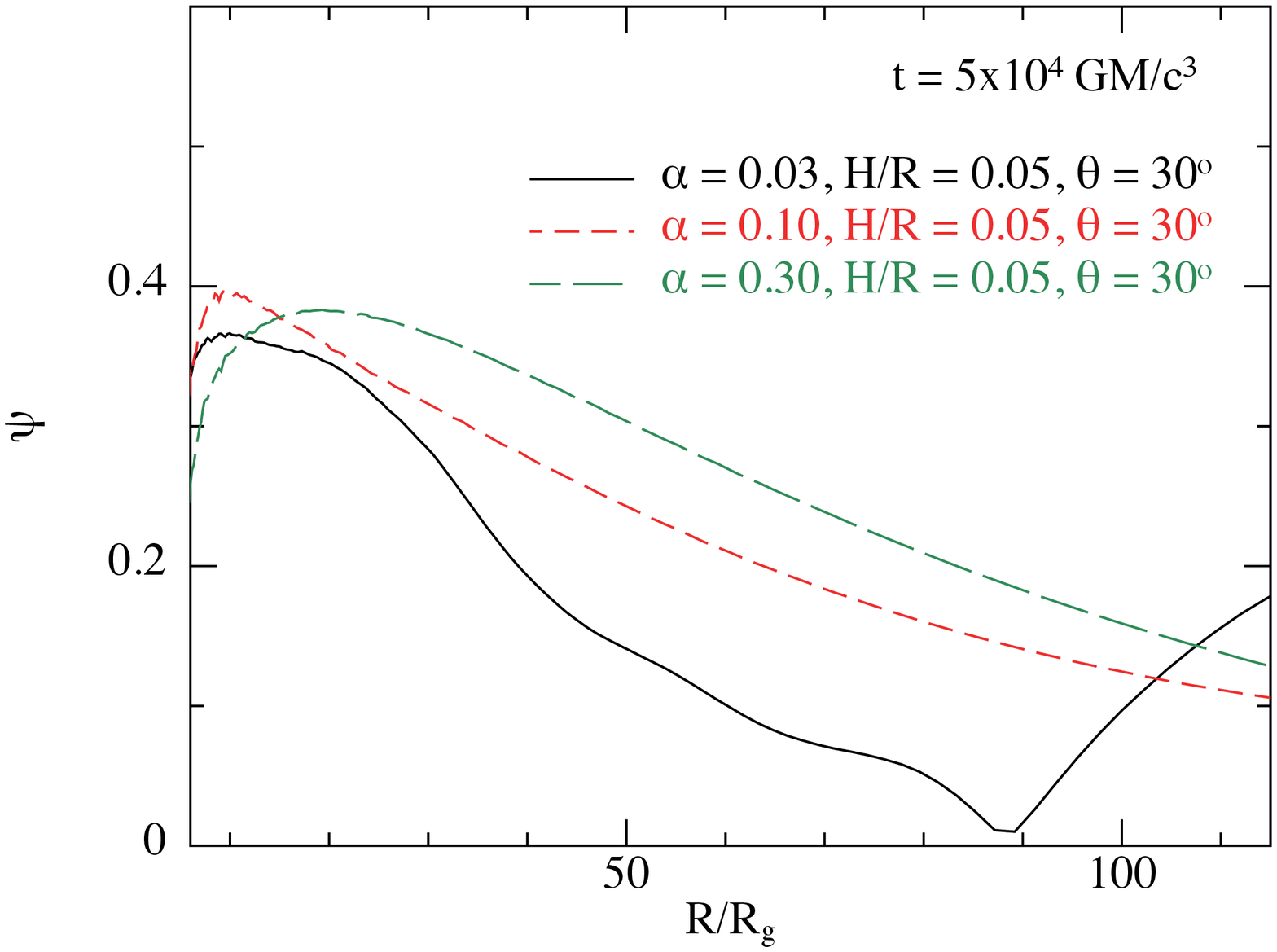}
  \caption{The warp amplitude, $\psi = R\left|\partial\mathbi{l}/\partial R\right|$, plotted against radius, $R/R_{\rm g}$, at a time of $5\times10^4 GM/c^3$ for the simulations with $\theta=30^\circ$. Each panel compares the warp amplitude in the simulations with varying values of the Shakura-Sunyaev viscosity parameter of $\alpha = 0.03$, $0.1$ and $0.3$. The left hand panel shows the results for the thinner case of $H/R=0.02$ and the right hand case shows the thicker case of $H/R=0.05$. The left hand panel shows that decreasing the viscosity can destabilise the disc, with the peak warp amplitude increasing until it becomes larger than the critical value. The warp amplitude at $R \approx 30R_{\rm g}$ in the higher viscosity cases is $\approx 0.5-1$. For $\alpha = 0.03$, the critical warp amplitude $\psi_{\rm c}\approx 0.5$, and thus the lower viscosity case becomes unstable when $\psi \gtrsim \psi_{\rm c}$ as expected. The right hand panel shows that for stable discs, as the viscosity is decreased the solutions can transition from diffusive to wavelike in character \cite[as shown by][]{Papaloizou:1983aa}.}
  \label{Fig5}
\end{figure*}

In Fig.~\ref{Fig5} we show the effect of varying the value of $\alpha$ in the simulations. In the left hand panel we show the simulations with $H/R = 0.02$ and $\theta=30^\circ$ for $\alpha = 0.03$, $0.1$ and $0.3$. In this case we can clearly see the non-linear effect of changing the disc viscosity on the disc warp. For the highest value of $\alpha$ the disc exhibits a smooth warped shape with a peak at $R/R_{\rm g} \approx 30$. As $\alpha$ is decreased the general shape across most of the disc remains the same, and the peak warp amplitude occurs in a similar radial location. However, for lower $\alpha$ the peak warp amplitude grows. This is a modest increase when $\alpha$ varies from $0.3$ to $0.1$, but reducing further to $\alpha=0.03$ renders the disc unstable and the peak warp amplitude increases sharply because of this.  In the right hand panel we show the simulations with $H/R = 0.05$ and $\theta =30^\circ$ for the same range of $\alpha$. In this case none of the simulations reach the critical warp amplitude and they all remain stable. The peak warp amplitude moves inwards slightly when $\alpha$ is decreased from $0.3$ to $0.1$ \citep[see also, for example,][]{Kumar:1985aa,Natarajan:1998aa}. These simulation results confirm the general expectation from \cite{Dogan:2018aa} that lower $\alpha$ is more likely to lead to instability. It is worth reminding the reader here that the values of viscosity here are the values for the physical viscosity term in the simulations, and that the total viscosity present also includes the numerical viscosity that we discussed in detail in Section~\ref{numcon}.

It is also worth noting that we detect the presence of propagating waves in the simulations with $\alpha=0.03$ and $H/R=0.05$ (i.e. with $\alpha \lesssim H/R$; see the low amplitude waves in the warp amplitude at $30 \lesssim R/R_{\rm g} \lesssim 80$ and the upturn in warp amplitude at $R/R_{\rm g} \gtrsim 90$ in the right hand panel of Fig.~\ref{Fig5}), but that these effects are noticeably reduced by the time $\alpha \gtrsim H/R$ (e.g. comparing the results with $\alpha = 0.03$ and $H/R = 0.02$ to those with $\alpha = 0.1$ and $H/R = 0.05$). This suggests that the transition from wavelike to diffusive as $\alpha$ is increased from below to above $H/R$ may be quite sharply focussed around $\alpha = H/R$ \citep[cf.][]{Martin:2019ab}. This is an interesting question that warrants a focussed investigation in the future. 

\subsection{Disc tearing}
In this sub-section we take a more detailed look at the properties of the simulated discs in which instability occurs. The simulations which exhibit clear disc tearing, and thus which we analyse in this section are the following:
\begin{enumerate}[label=\Alph*:]
\item $\alpha = 0.03$, $H/R=0.02$, $\theta = 30^\circ$,
\item $\alpha=0.03$, $H/R=0.02$, $\theta=60^\circ$,
\item $\alpha = 0.03$, $H/R=0.05$, $\theta = 60^\circ$, and
\item $\alpha=0.1$, $H/R=0.02$, $\theta=60^\circ$.
\end{enumerate}
For these simulations we note that B is the same as A, but at higher inclination. C is the same as B, but with a thicker disc. And D is the same as B, but with a higher viscosity. One of the other simulations, with $\alpha = 0.03$, $H/R = 0.05$, $\theta=30^\circ$, shows an early break, which subsequently decays after propagating outwards through the disc, so we do not include this in the analysis. We also find instability in the case of
\begin{enumerate}[label=\Alph*:,start=5]
\item $\alpha = 0.3$, $H/R = 0.02$, $\theta=60^\circ$.
\end{enumerate}
As this case is anomalous from the point of view of the local stability analysis presented by \cite{Dogan:2018aa} we discuss it separately in Section~\ref{alpha0.3} below. The remaining simulations all warp in response to Lense-Thirring precession, but do not exhibit any instability.

We discuss in turn each of the simulations that show clear examples of disc tearing behaviour (listed A-E above). In each case we identify two regions of the disc which exhibit a disc break caused by disc tearing. We look for clean examples where the disc is coherent beforehand and results in a single clear break (at least locally to that patch of the disc). For example, we have not chosen regions where a peak in warp amplitude ultimately splits into several distinct breaks, or where the break occurs in a region where there is already a complex shape to the warp amplitude. In such regions we expect the effects of different growing modes would complicate matters and make our analysis less clear. For the two breaks examined, in each case we measure the radial range ($R_1,R_2$) within which the peak of the warp amplitude resides while the peak warp amplitude grows from $\psi_0$ at time $t_0$ to $\psi_1$ at time $t_1$.\footnote{Note that when searching for the peak warp amplitude we find that the 150 logarithmically spaced radial points used in the previous sections is insufficient. For these calculations we employ a linear spaced grid with 1200 points.} We then calculate the time evolution of the peak warp amplitude in this region $\psi_{\rm p}(t)$. In unstable regions we expect the warp amplitude to grow as
\begin{equation}
  \label{psit}
  \psi(t) \sim \psi_0\exp\left[\Re\{s\}\Omega\left(Hk\right)^2(t-t_0)\right]\,,
\end{equation}
where $\Omega$ is the orbital frequency, $H\equiv c_{\rm s}/\Omega$, $k$ is the wavenumber, $\Re\{s\}$ is the dimensionless growth rate of the instability. In each unstable case, we find that the evolution of the peak warp amplitude with time is approximately exponential while the peak warp amplitude is in the range $2.5-4$ (in some cases the exponential growth starts at lower warp amplitude and in some cases it continues to larger warp amplitudes, but we find this range can be consistently applied to most of our cases; when we use a different range we note it below). To estimate the growth rate, we rearrange (\ref{psit}) to give
\begin{equation}
 \Re\{s\}(Hk)^2 = \frac{\ln(\psi_1/\psi_0)}{\Omega \Delta t}\,,
\end{equation}
where $\Delta t = t_1-t_0$ is the time taken to grow from $\psi_0 = 2.5$ to $\psi_1 = 4$. We find that in each case the peak of the warp amplitude does not move significantly while the amplitude grows between these two bounds, allowing us to use a single value for $\Omega$, which for the Einstein potential we employ is given by
\begin{equation}
  \Omega^2 = \frac{GM}{R^3}\left(1 + \frac{6R_{\rm g}}{R}\right)\,. 
\end{equation}

Ideally, we would also like to measure the wavenumber, $k$, from the simulations so that we could compare the value of the dimensionless growth rate directly to the predicted value. However, there does not appear to be an accurate way to measure this directly from the simulation data. Future simulations which insert a growing mode with a single, well-defined value for $k$ are desirable. In the dynamic simulations we present here it is likely that there is significant power at a range of wavenumbers. In this case we can expect growth to be slower initially, until sufficient power is in the fastest growing modes. For the mode to fit in the disc (recall that we cannot bend the disc on scales $\lesssim H$) we must have $Hk \lesssim 1$ \citep{Ogilvie:2000aa}, so we can expect the fastest growing modes appear with $Hk$ less than, but not much less than, unity. However, to be consistent we report the values of $\chi \equiv \Re\{s\}(Hk)^2$ which can be measured from the simulation data.

For simulation A with $\alpha = 0.03$, $H/R = 0.02$ and $\theta = 30^\circ$ we find that the warp amplitude in the inner disc regions rapidly grows to greater than the critical value for $R\lesssim 15 R_{\rm g}$ (which for this viscosity is $\psi_{\rm c} \approx 0.5$). This inner region breaks into multiple rings, starting from the inner parts and ending at $\approx 30 R_{\rm g}$. Outside of this region $\psi$ remains below the critical value at all times, and no breaks are observed in this region. For this simulation we analyse two clear breaks, the first occurring between $R\approx 12-18 R_{\rm g}$ with a peak in the warp amplitude at $R_{\rm p} \approx 15R_{\rm g}$ while the warp amplitude is growing from $\psi_0 = 2.5$ to $\psi_1 = 4$. The second break forms an initial peak in the warp amplitude at $R\approx 20R_{\rm g}$ and by the time the peak warp amplitude is $\approx 2$ the peak has settled at a radius of $R_{\rm p}\approx 27 R_{\rm g}$. For the first break, we measure $\Delta t \approx 1500$ (in units of $GM/c^3$) and with the peak at $R_{\rm p}\approx 15R_{\rm g}$ we have $\Omega \approx 0.02$. This yields $\chi \approx 0.015$. For the second peak we have $\Delta t \approx 4000$, $R_{\rm p}\approx 27R_{\rm g}$, $\Omega \approx 8\times10^{-3}$ and therefore $\chi \approx 0.015$. This means that, when scaled to the local dynamical time, the growth rate was approximately the same in both cases as expected. In this simulation we expect the dimensionless growth rate to be in the range of $0.2-0.5$ (cf. Figs~\ref{Fig1} \& \ref{Fig2}). If $\Re\{s\} = 0.2-0.5$, this would imply $k^{-1} \approx (3-5)H$ which seems reasonable, but we also note that it is likely that the total level of viscosity in the simulation is slightly higher than $\alpha=0.03$ (see discussion in Section~\ref{numcon}) and thus we would expect the growth rate to be reduced somewhat (i.e. if $\alpha \approx 0.05$, then the predicted dimensionless growth rate implies $k^{-1} \approx (2-3)H$).

For simulation B with $\alpha = 0.03$, $H/R = 0.02$ and $\theta = 60^\circ$ we analyse two breaks in the disc, with the first occurring at $R_{\rm p}\approx 24 R_{\rm g}$ and the second which begins at $R\approx 30R_{\rm g}$ and settles into a strong peak at $R_{\rm p}\approx 40 R_{\rm g}$. For the first break, we measure $\Delta t \approx 1650$ and with the peak at $R_{\rm p}\approx 24R_{\rm g}$ we have $\Omega \approx 0.01$. This yields $\chi \approx 0.03$. For the second peak we have $\Delta t \approx 5400$, $R_{\rm p}\approx 40R_{\rm g}$, $\Omega \approx 4\times10^{-3}$ and therefore $\chi \approx 0.02$. As before, when scaled to the local dynamical timescale the growth rate in the unstable region is similar between the two breaks in the same disc, and are also similar to the previous simulation which has the same $\alpha$ value. 

For simulation C with $\alpha = 0.03$, $H/R = 0.05$ and $\theta = 60^\circ$ we also analyse two breaks in the disc, with the first occurring between $R\approx 20-30 R_{\rm g}$ with the peak typically at $R_{\rm p}\approx 25 R_{\rm g}$ and the second occurring between $R\approx 30-50 R_{\rm g}$ with the peak typically at $R_{\rm p}\approx 45 R_{\rm g}$. For the first break, we measure $\Delta t \approx 2400$ and with the peak at $R_{\rm p}\approx 25R_{\rm g}$ we have $\Omega \approx 9\times 10^{-3}$. This yields $\chi \approx 0.02$. For the second peak we note that the growth in warp amplitude slows appreciably for $\psi \gtrsim 3.5$, but is approximately exponential between $\psi = 2.5$ and $\psi = 3.5$, so in this instance we explore this range. Here we have $\Delta t \approx 4000$, $R_{\rm p}\approx 45R_{\rm g}$, $\Omega \approx 3.5\times10^{-3}$ and therefore $\chi \approx 0.025$. These values are similar to the previous two simulations. It is worth noting that this simulation is of a thicker disc with $H/R = 0.05$, and in this case the radial extent of the peak of the warp amplitude is broader than in the thinner disc simulations by a factor of approximately 2.5 which demonstrates that the break in the disc occurs across a radial length scale of a few $H$ independent of the value of $H/R$. 

For simulation D with $\alpha = 0.1$, $H/R = 0.02$ and $\theta = 60^\circ$ we again analyse two breaks in the disc, with the first occurring at $R_{\rm p}\approx 22 R_{\rm g}$ and the second occurring at $R_{\rm p}\approx 37R_{\rm g}$. For the first break, we measure $\Delta t \approx 1500$ and with the peak at $R_{\rm p}\approx 22R_{\rm g}$ we have $\Omega \approx 0.01$. This yields $\chi \approx 0.03$. For the second peak we have $\Delta t \approx 5000$, $R_{\rm p}\approx 37R_{\rm g}$, $\Omega \approx 5\times10^{-3}$ and therefore $\chi \approx 0.02$. Therefore we find that the growth rates in this case, with $\alpha=0.1$, are similar to the growth rates found for $\alpha=0.03$. The predictions from the stability analysis suggest that the growth should be slower in this case. This may indicate that for the lower viscosity simulations ($\alpha=0.03$) the numerical viscosity is impeding the growth rate somewhat. Alternatively it might be that initially growth occurs for small $k$ (on large spatial scales) and thus the growth is slow until enough power is transferred into large $k$ modes (smaller length scales) at which point the instability is sufficiently rapid that the timescale is essentially independent of disc parameters once the disc becomes unstable. We note that $\chi\approx 0.02$ implies that the instability growth timescale is $\sim 50/\Omega$ (i.e., several orbits). It seems unreasonable to expect the instability to grow faster than this. We speculate that this second reason is what is occurring in the simulations as we find that analysis of two of the breaks in the simulation with $\alpha = 0.1$, $H/R = 0.02$ and $\theta = 60^\circ$ when performed at a resolution corresponding to $N_{\rm p} = 10^6$ both also yield $\chi \approx 0.02$, hinting that the growth rate (which encompasses the wavenumber of the fastest growing mode) may not depend strongly on resolution. 

Finally, we also note that simulation C, with $\alpha = 0.03$, $H/R = 0.05$ and $\theta = 60^\circ$, performed at a resolution corresponding to $N_{\rm p} = 10^6$ showed some interesting dynamics that was not present in any of the other simulations. In this simulation the peak of the warp amplitude is concentrated at small radii. There are two additional peaks of warp amplitude, which occur at larger radii. In the $N_{\rm p} = 10^7$ simulation, these peaks at larger radii are resolved into breaks which then alter the inner disc evolution, while in the $N_{\rm p} = 10^6$ case the outer disc remained coherent. This allowed the inner regions, on longer timescales, to develop a break near the inner edge which is quickly accreted. This process repeats with the innermost ring being broken off, it subsequently precesses, and it is then accreted. It is possible that for the right combination of parameters, and variation of those parameters with radius, that this evolution could manifest in a physical disc. For example, if the disc angular semi-thickness were a slowly increasing function of radius, then the disc may only be unstable near the inner disc edge (and this could remain so as resolution is increased). We discuss this manifestation of the disc tearing dynamics, and the implications this may have for the observational properties of these discs, in more detail in \cite{Raj:2021ab}.

\subsection{Instability at high viscosity}
\label{alpha0.3}
We noted above that simulation E, with $\alpha = 0.3$, $H/R = 0.02$ and $\theta = 60^\circ$, showed instability. This was not expected as the local stability analysis of the warped disc equations suggests that the disc is stable for any value of the warp amplitude at this high a viscosity. However, instability in this case had previously been reported by \cite{Nixon:2012aa} for discs subject to the Lense-Thirring precession using a 1D ring code to evolve the warped disc evolution equations. In this case \cite{Nixon:2012aa} showed that the disc can break into two distinct planes for $\alpha \approx 0.2-0.3$. They present evolution that is distinct from the evolution presented by most hydrodynamical simulations in that the breaks manifest as a sharp variation in the disc tilt, while the breaks in most hydrodynamical simulations manifest (at least initially) as a sharp variation in the disc twist. Here we find that for $\alpha=0.3$, the break in the disc forms with a large variation in the disc tilt, consistent with the results in \cite{Nixon:2012aa}.

For comparison with the growth rates reported above, we again analyse two breaks in the disc for this simulation with $\alpha=0.3$. The first occurring at $R_{\rm p}\approx 14 R_{\rm g}$ and the second occurring at $R_{\rm p}\approx 18R_{\rm g}$. For the first break, we measure $\Delta t \approx 950$ and with the peak at $R_{\rm p}\approx 14R_{\rm g}$ we have $\Omega \approx 0.02$. This yields $\chi \approx 0.02$. For the second peak we have $\Delta t \approx 1250$, $R_{\rm p}\approx 18R_{\rm g}$, $\Omega \approx 0.015$ and therefore $\chi \approx 0.025$. Therefore we find that the growth rates in this case, with $\alpha=0.3$, are similar to the growth rates found for the lower $\alpha$ cases.

There are several possibilities for why the simulated disc with $\alpha=0.3$ may be unstable, contrary to the prediction of the local stability analysis. It is possible, although we consider it unlikely, that there exists an island of instability at high $\alpha$ for some values of the shear rate $q$ and bulk viscosity $\alpha_{\rm b}$ that was not revealed in the analyses presented in \cite{Dogan:2018aa} and \cite{Dogan:2020aa}. It is possible that the disc tearing presented in these simulations is a numerical artefact, but again we consider this unlikely given the broad range of codes and methods that have now reported such behaviour in strongly warped discs. We speculate that the answer is related to the applicability of the warped disc equations (and in particular the accuracy of calculating the torque coefficients) and/or the local stability analysis at the extremes of the parameter space. Here we have large viscosity and large warp amplitudes, and the disc structure is strongly affected by the external torque. It seems possible that one or more of the assumptions of the theory of warped discs could be violated, or that non-local effects become important. For example, as discussed by \cite{Nixon:2012aa}, the torques responsible for holding the disc together generally weaken with increasing warp amplitude. This is in part due to the drop in surface density in the region of large warp amplitude. Once the surface density has dropped sufficiently far, it may be that there is simply not a strong enough torque left to hold the disc together in that region. If the stability of the disc is dependent on such a process, in which the local mass is slowly reduced over time, this may not be captured by a local stability analysis which assumes a stable background state.

\section{Conclusions}
\label{conclusions}
We have presented and discussed the results of numerical simulations of discs that are warped by the Lense-Thirring effect from a spinning black hole. In general we find that the results of our simulations (Section~\ref{sims}) are consistent with the predictions of the local stability analysis of the warped disc equations \citep{Dogan:2018aa}. We confirm that disc tearing occurs preferentially in thin, low viscosity, and highly inclined discs. Thin and highly inclined discs are more likely to yield large warp amplitudes, and low viscosity discs are generally expected to become unstable at lower warp amplitudes. We show that instability arises in the simulated discs at higher values of the warp amplitude for larger viscosity parameters, and the critical warp amplitudes are commensurate with the predicted values. We also find that the growth rates of the instability are consistent with the predicted growth rates, assuming that the growing modes have wavenumbers, $k$, with $kH \sim 0.3-0.5$. We also find some unexpected results. In particular we found instability for a simulation with $\alpha=0.3$ (case E). In this case the instability manifests with a sharp change in the disc tilt, rather than the disc twist, as previously found for similar high viscosity cases by \cite{Nixon:2012aa} using a 1D ring code approach. This suggests that there is a difference between the dynamics in numerical simulations that include a strong external torque to provide the precession of disc orbits and the dynamics predicted by a local stability analysis of the warped discs equations that do not include the external torque, with which \cite{Dogan:2018aa} find that such large $\alpha$ value should be stable. We note that we do find that the results of the simulations vary somewhat with increasing resolution; typically we find that for stable discs the results are converged for the bulk of the disc with small differences near the inner edge where the density, and thus resolution, is lower, but we do find that higher resolution simulations exhibit more disc breaks that are sharper (i.e.\@ have a larger peak warp amplitude; see Appendix~\ref{res}). We attribute this to the increase in spatial resolution which allows smaller physical features to be resolved in the simulations, and to the decreasing influence of numerical viscosity at higher resolution. Returning to the three aims that we listed at the end of Section~\ref{numcon}, namely
\begin{enumerate}
\item discs with warp amplitude above the critical warp amplitude, $\psi > \psi_{\rm c}$, for extended periods of time lead to instability,
\item discs with $\psi < \psi_{\rm c}$ are stable, and
\item that the growth rate of the warp amplitude in the unstable regions follow the general trends of the predicted growth rates, i.e. that the growth rates are generally higher for smaller viscosity and depend on the warp amplitude.
\end{enumerate}
Our simulations support (1) and (2), showing that the instability seen in the simulations is consistent with the predicted critical warp amplitudes. However, for (3), while the growth rates are consistent at the order of magnitude level, future simulations at higher resolution will be required to determine if the growth rates vary in the predicted manner with disc parameters such as the viscosity parameter and warp amplitude. We discuss the observational implications of the disc tearing behaviour for black hole accretion in \cite{Raj:2021ab}.

\acknowledgments
We thank the referee for a helpful report. We thank Jim Pringle for useful comments on the manuscript. CJN is supported by the Science and Technology Facilities Council (grant number ST/M005917/1). CJN acknowledges funding from the European Union’s Horizon 2020 research and innovation program under the Marie Sk\l{}odowska-Curie grant agreement No 823823 (Dustbusters RISE project). This research used the ALICE High Performance Computing Facility at the University of Leicester. This work was performed using the DiRAC Data Intensive service at Leicester, operated by the University of Leicester IT Services, which forms part of the STFC DiRAC HPC Facility (\url{www.dirac.ac.uk}). The equipment was funded by BEIS capital funding via STFC capital grants ST/K000373/1 and ST/R002363/1 and STFC DiRAC Operations grant ST/R001014/1. DiRAC is part of the National e-Infrastructure. We used {\sc splash} \citep{Price:2007aa} for the figures.

\bibliographystyle{aasjournal}
\bibliography{nixon}
%%%%%%%%%%%%%%%%% APPENDICES %%%%%%%%%%%%%%%%%%%%%
\appendix
\section{Resolution}
\label{res}
\begin{figure*}
  \centering
  \includegraphics[width=0.33\textwidth]{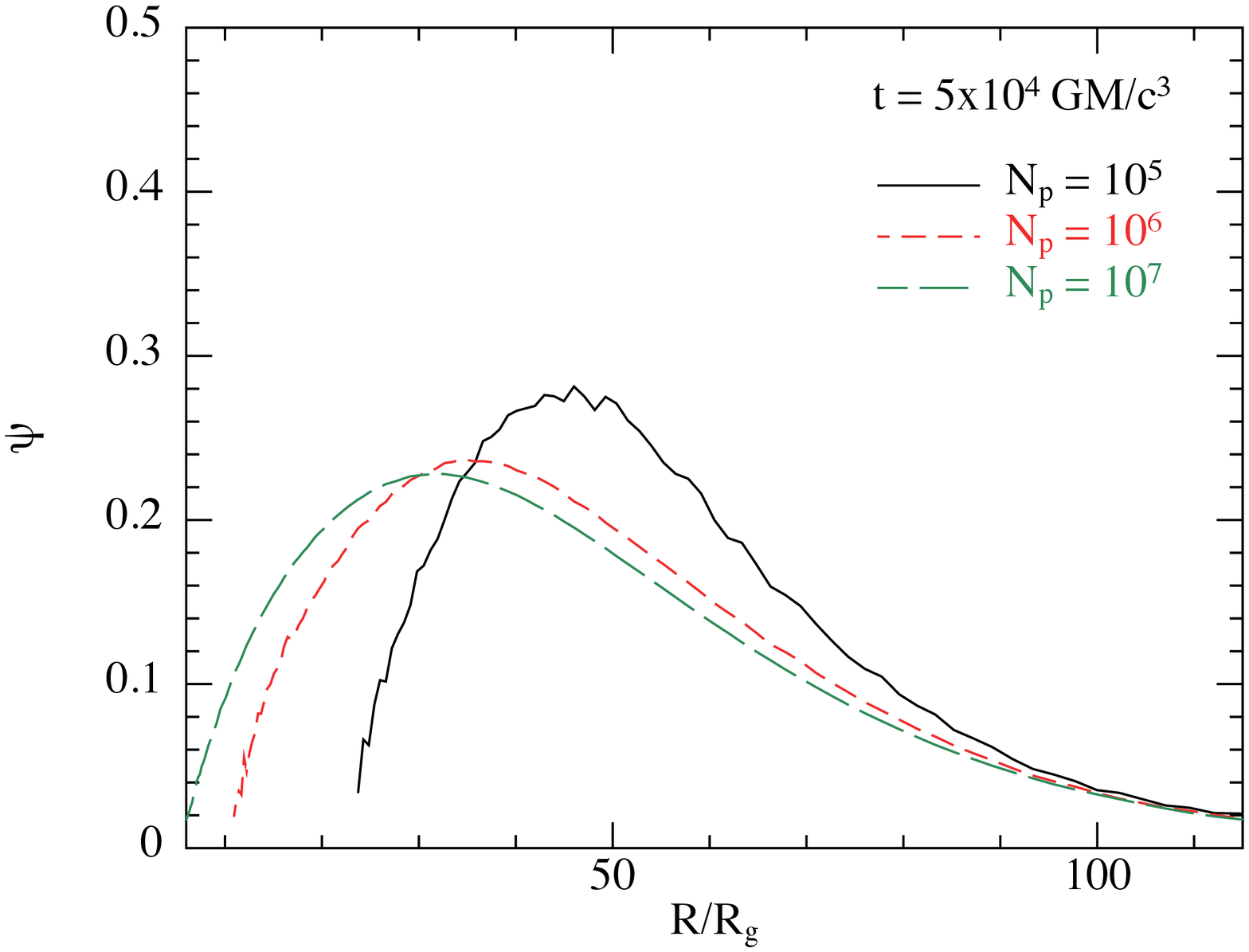}\hfill
  \includegraphics[width=0.33\textwidth]{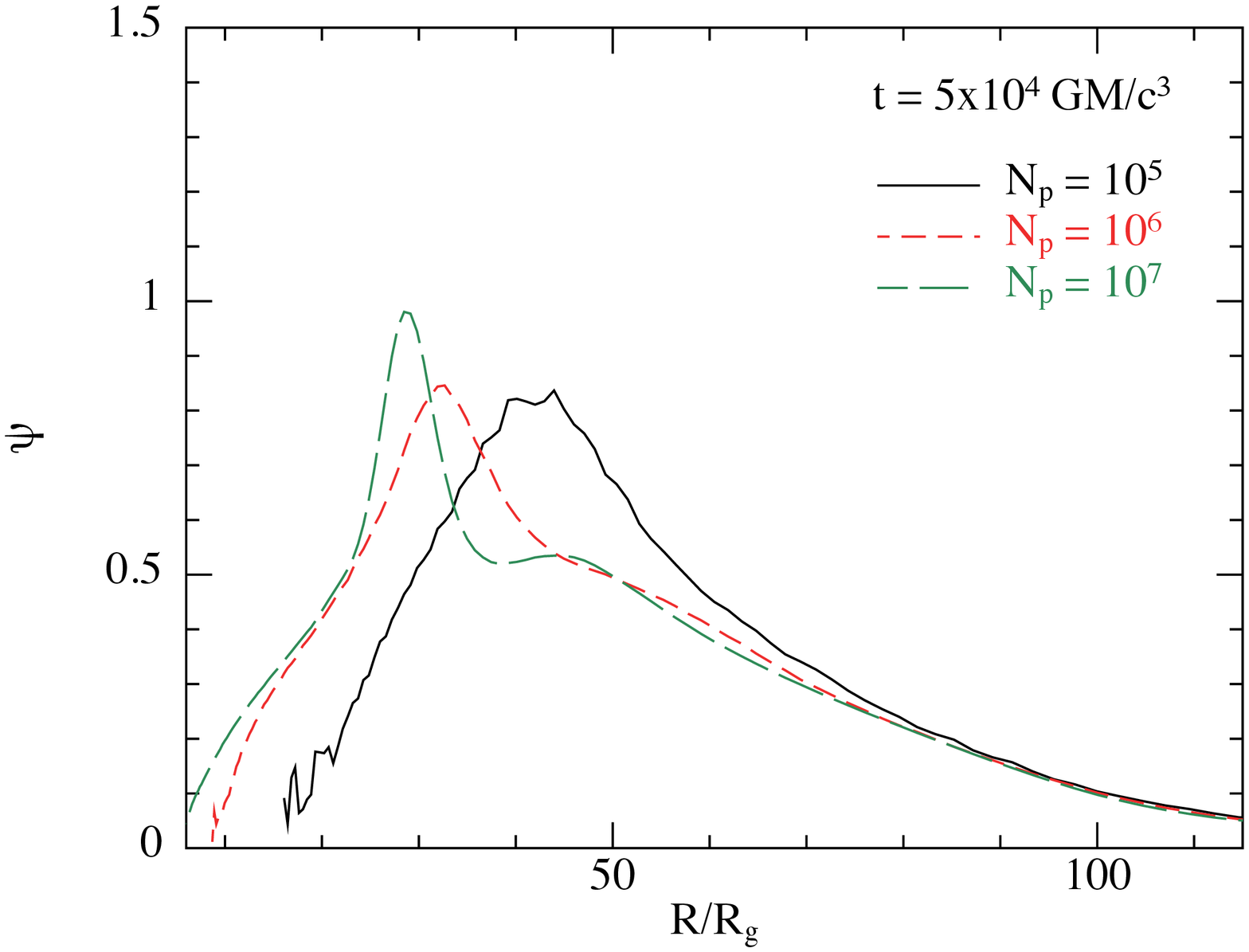}\hfill
  \includegraphics[width=0.33\textwidth]{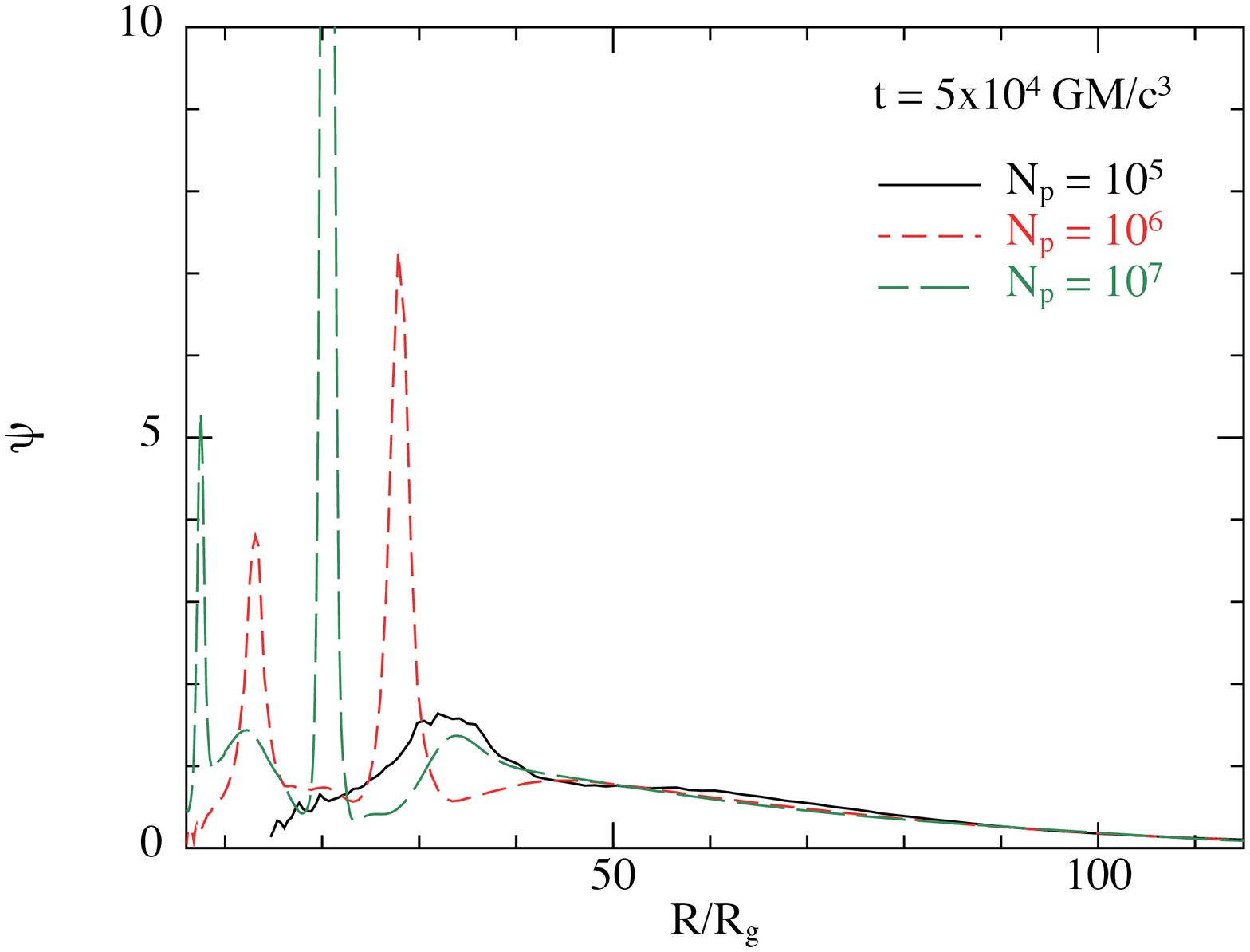}
  \caption{The warp amplitude, $\psi = R\left|\partial\mathbi{l}/\partial R\right|$, plotted against radius, $R/R_{\rm g}$, at a time of $5\times10^4 GM/c^3$ for the simulations at different inclinations (left to right) with $\alpha = 0.1$ and $H/R = 0.02$. Each panel compares the warp amplitude in the simulations varying the resolution as given in the legend. The left hand panel shows the results for the inclination of $10^\circ$, the middle panel shows the inclination of $30^\circ$ and the right hand panel shows the inclination of $60^\circ$. At low inclinations the disc is stable, and the results are similar at high particle number. The same is true at inclination of $30^\circ$. At $60^\circ$ the disc is unstable for these parameters, and the solutions all exhibit this but the details are different at different resolutions as discussed in the text.}
  \label{FigX}
\end{figure*}

We have discussed the effects of resolution in detail in Section~\ref{numcon}, and throughout. There are two principle effects that numerical resolution has on the results of warped disc simulations. First, higher spatial resolution allows for the formation of sharper features in the disc. For example, an underresolved region of high warp amplitude may result in the warp amplitude being underestimated and thus the disc remaining (artificially) stable when (physically) the disc should be unstable. Secondly, the numerical viscosity in SPH simulations is a function of the local resolution with the linear numerical viscosity proportional to the resolution lengthscale and the quadratic numerical viscosity proportional to the square of the resolution lengthscale (cf. equation~\ref{alphaAV}). As the resolution is increased the contribution to the total viscosity from the numerical viscosity decreases, and thus for a fixed physical viscosity the total viscosity in the simulation decreases. As discussed in Section~\ref{numcon} we expect the magnitude of the numerical viscosity to be of the order of, but smaller than, the lowest physical viscosity we simulated ($\alpha = 0.03$). Therefore we expect the simulations with lower physical viscosity to be more strongly affected by it than the higher viscosity simulations with say $\alpha=0.3$. Both of these effects combine at the inner edge of the disc in SPH simulations where the surface density goes to zero at the inner boundary. Here, spatial gradients in the disc (for example of the surface density) are typically largest and may be underresolved by the available spatial resolution, and the decreasing spatial resolution associated with the decreasing density of the disc results in an increased viscosity. As resolution is increased the solution tends towards the expected solution (e.g. equation~\ref{sigr} for the surface density of a planar disc). Finally, we also noted that because the instability, including both the critical warp amplitude and the growth rates \citep[see][for details]{Dogan:2018aa}, depends sensitively on the viscosity parameter $\alpha$, we cannot expect numerical simulations to show converged results for the precise details of unstable regions, for example the exact location at which the instability manifests. To show an example of the differences in the simulation results at different resolutions we plot the warp amplitudes for the simulations at different inclination angles with $\alpha=0.1$ and $H/R=0.02$ in Fig.~\ref{FigX} each time for different numbers of particles.

%%%%%%%%%%%%%%%%%%%%%%%%%%%%%%%%%%%%%%%%%%%%%%%%%%

\end{document}